\documentclass[a4paper,fleqn,usenatbib]{mnras}

\usepackage{mathptmx}
\usepackage[T1]{fontenc}
\usepackage{ae,aecompl}

\usepackage{graphicx}	% Including figure files
\usepackage{amsmath}	% Advanced maths commands
\usepackage{amssymb}	% Extra maths symbols
\usepackage{tikz}
\usepackage{multirow}
\usepackage{upgreek}
\usepackage{accents}

\def\kms{\,km\,s$^{-1}$}
\def\msun{\,M$_{\odot}$}
\newcommand*{\dt}[1]{\accentset{\mbox{\large\bfseries .}}{#1}}
\newcommand{\ltsimeq}{\raisebox{-0.6ex}{$\,\stackrel{\raisebox{-.2ex}{$\textstyle <$}}{\sim}\,$}}
\newcommand{\gtsimeq}{\raisebox{-0.6ex}{$\,\stackrel{\raisebox{-.2ex}{$\textstyle >$}}{\sim}\,$}}
\newcommand{\farc}{\mbox{\ensuremath{.\!\!^{\prime\prime}}}}

\title[WISDOM project -- VIII. Multi-scale feedback in NGC\,0708]{WISDOM project -- VIII. Multi-scale feedback cycles in the brightest cluster galaxy NGC\,0708}
\author[E. V. North et al.]{\parbox{\textwidth}{
		Eve V. North,$^{1}$
		Timothy A. Davis,$^{1}$\thanks{E-mail: DavisT@cardiff.ac.uk}
		Martin Bureau,$^{2,3}$
		Massimo Gaspari,$^{4,5}$
		Michele Cappellari,$^{2}$
		Satoru Iguchi,$^{6,7}$
		Lijie Liu,$^{2}$
		Kyoko Onishi,$^{8,9}$
		Marc Sarzi,$^{10}$
		Mark D. Smith$^{2}$
		and Thomas G. Williams$^{1}$}
	    \vspace{0.4cm}
\\
\parbox{\textwidth}{
$^{1}$School of Physics \& Astronomy, Cardiff University, Queens Buildings, The Parade, Cardiff CF24 3AA, UK\\
$^{2}$Sub-department of Astrophysics, Department of Physics, University of Oxford, Denys Wilkinson Building, Keble Road, Oxford OX1 3RH, UK\\
$^{3}$Yonsei Frontier Lab and Department of Astronomy, Yonsei University, 50 Yonsei-ro, Seodaemun-gu, Seoul 03722, Republic of Korea\\
$^{4}$INAF, Osservatorio di Astrofisica e Scienza dello Spazio, via P. Gobetti 93/3, 40129 Bologna, Italy\\
$^{5}$Department of Astrophysical Sciences, Princeton University, Princeton, NJ 08544, USA \\
$^{6}$Department of Astronomical Science, SOKENDAI (The Graduate University of Advanced Studies), Mitaka, Tokyo 181-8588, Japan\\
$^{7}$National Astronomical Observatory of Japan, National Institutes of Natural Sciences, Mitaka, Tokyo 181-8588, Japan\\
$^{8}$Research Center for Space and Cosmic Evolution, Ehime University, Matsuyama, Ehime, 790-8577, Japan\\
$^{9}$Department of Space, Earth and Environment, Chalmers University of Technology, Onsala Observatory, SE-439 92 Onsala, Sweden\\
$^{10}$Armagh Observatory and Planetarium, College Hill, Armagh, BT61 9DG, UK\\
$^{11}$Max Planck Institut f\"{u}r Astronomie, K\"{o}nigstuhl 17, 69117 Heidelberg, Germany\\}
}
\date{Accepted 2021 March 15. Received 2021 March 15; in original form 2020 July 21}

\pubyear{2021}

\begin{document}
\label{firstpage}
\pagerange{\pageref{firstpage}--\pageref{lastpage}}
\maketitle

\begin{abstract}
We present high-resolution (synthesised beam size 0\farcs088$\times$0\farcs083 or $25\times23$\,pc$^{2}$) Atacama Large Millimetre/submillimetre Array (ALMA) $^{12}$CO(2--1) line and 236\,GHz continuum observations, as well as 5\,GHz enhanced Multi-Element Radio Linked Interferometer Network  (e-MERLIN) continuum observations, of NGC\,0708; the brightest galaxy in the low-mass galaxy cluster Abell\,262. The line observations reveal a turbulent, rotating disc of molecular gas in the core of the galaxy, and a high-velocity, blue-shifted feature $\approx0$\farcs4 ($\approx113$\,pc) from its centre. The sub-millimetre continuum emission peaks at the nucleus, but extends towards this anomalous CO emission feature. No corresponding elongation is found on the same spatial scales at 5\,GHz with e-MERLIN. We discuss potential causes for the anomalous blue-shifted emission detected in this source, and conclude that it is most likely to be a low-mass in-falling filament of material condensing from the hot intra-cluster medium via chaotic cold accretion, but it is also possible that it is a jet-driven molecular outflow. We estimate the physical properties this structure has in these two scenarios, and show that either explanation is viable. We suggest future observations with integral field spectrographs will be able to determine the true cause of this anomalous emission, and provide further evidence for interaction between quenched cooling flows and mechanical feedback on both small and large scales in this source.
\end{abstract}
\begin{keywords}
galaxies: individual: NGC\,0708, galaxies: active , ISM: jets and outflows, galaxies: elliptical and lenticular, cD- , galaxies: clusters: intracluster medium, galaxies: kinematics and dynamics
\end{keywords}

%%%%%%%%%%%%%%%%%%%%%%%%%%%%%%%%%%%%%%%%%%%%%%%%%%

%%%%%%%%%%%%%%%%% BODY OF PAPER %%%%%%%%%%%%%%%%%%

\section{Introduction}

The lack of molecular gas in early-type galaxies (ETGs; ellipticals and S0s) has been a point of debate for some decades (e.g. \citealt{Faber1976}; \citealt{Lees1991}; \citealt{Young2011}; \citealt{Davis2019}). Observations show that whilst ETGs have internal sources of gas, for instance stellar mass-loss, they have lower gas fractions than late-type galaxies \citep[e.g.][]{Lees1991}. This is especially true of brightest cluster galaxies (BCGs), where mergers and intra-cluster medium (ICM) cooling should bring large additional amounts of molecular gas into the galaxy, but their observed molecular gas reservoirs are an order of magnitude smaller than expected (based on cooling rates estimated from X-ray emission; e.g. \citealt{Lees1991}; \citealt{Fabian1994}; \citealt{DeLucia2007}). 

{Searches for this cooled gas have persistently returned lower gas masses and fewer young stars than required by cooling flow observations (e.g. \citealt{Johnstone1987}; \citealt{Heckman1989}; \citealt{McNamara1989}; \citealt{Crawford1999}; \citealt{Donahue2000}; \citealt{Hoffer2012}). For instance, in a range of brightest cluster galaxies the mass of molecular gas is found to be $5-10$\,per\,cent of that expected from the cooling flow (i.e. \citealt{Edge2001}; \citealt{Salome2004}). A solution to this `cooling flow problem', how gas leaves the hot phase but does not condense at the expected large rates (of 100--1000\msun\,yr$^{-1}$) on to the central galaxy, has been sought ever since.}

{High-angular resolution X-ray observations paved the way for answers, showing that despite appearing relaxed at low resolution, the centres of cooling flow clusters are in fact very dynamic. Active galactic nuclei (AGN) with powerful jets are found in essentially all cooling flow cluster central galaxies \citep{Sun2009}, and they are the principal power source driving the ICM dynamics (e.g. \citealt{Birzan2004,Birzan2012}; \citealt{McNamara2005}; \citealt{Rafferty2006}; \citealt{McNamara2007,McNamara2012}; \citealt{Gaspari2013}; \citealt{Hlavacek-Larrondo2015}). 
AGN jets appear able to inflate large bubbles in the hot ICM, that rise buoyantly and disrupt the cooling flows. 
Heat from the AGN is also distributed in the ICM through turbulent mixing and cocoon shocks (e.g. \citealt{Gaspari2013,2019ApJ...871....6Y,2020MNRAS.498.4983W}).
It has been shown that AGN jets have the mechanical power to balance the ICM's energy losses due to cooling, motivating the theory that mechanical (i.e. radio-mode) feedback is the principal regulator of ICM cooling, thus preventing a run away process (see reviews from e.g. \citealt{McNamara2007,McNamara2012,2020NatAs...4...10G}).} {Simulations also suggest AGN feedback is vital for the regulation of a galaxies gas reservoir and therefore its star formation rate. 
For instance, models including radio-mode feedback are in better agreement with e.g. the galaxy luminosity function (\citealt{Bower2006, Bower2008}; \citealt{Croton2006}; \citealt{McCarthy2008}; \citealt{Dave2012}).} 

{The advent of high-resolution radio/sub-mm interferometry has begun to add to the growing picture of feedback-controlled galaxy evolution. In multiple cooling flow clusters significant ($10^{9}$-$10^{10}$\msun) amounts of molecular gas have been detected in filaments which are sometimes coincident with buoyant X-ray bubbles rising through the ICM (e.g. \citealt{McNamara2014}; \citealt{Russell2014,Russell2016,Russell2017a,Russell2017b}; \citealt{Vantyghem2016,Vantyghem2018}), and sometimes present throughout the inner ICM (\citealt{Temi2018,2019MNRAS.489..349R,2019MNRAS.484.2886J}). It is not currently known if this cold gas has recently cooled from low-entropy gas lifted by the bubble, or is stimulated to cool in-situ by its passing. 
Many of these observed filaments also have star formation associated with them (e.g. \citealt{Vantyghem2018}).}

{A variety of works have used simulations to look at the formation of this multiphase ICM to ascertain how it is regulated. \citet{Gaspari2012}, \citet{Sharma2012}, \citet{Prasad2015} and \citet{Li2015} all see cycles within their simulations where dense, cold gas filaments condense out of the ICM and precipitate onto the central galaxy. This causes star formation and fuels the central SMBH (that becomes an AGN) via a process known as chaotic cold accretion (CCA; see \citealt{2019ApJ...884..169G} for a recent overview). The AGN and supernova winds increase heating within the ICM, returning it to a high entropy state, stopping cooling and hence the fuel supply. When the heating stops, cooling resumes again. They find that cold gas filaments form when the instantaneous ratio of the thermal instability and free-fall (or eddy turnover) timescales is $\ltsimeq10$ (e.g. \citealt{Gaspari2012}; \citealt{Sharma2012}, \citealt{Li2015}; \citealt{2018ApJ...854..167G}). 

{In all scenarios described above, the AGN is a crucial driver of the evolution of the gaseous material in brightest cluster galaxies, with jet-blown bubbles dominating on the scale of ten to hundreds of kilo-parsecs. The feedback produced by these AGN, however, interacts with multiphase condensation processes across a large range of physical scales. \citet{Tremblay2012a, Tremblay2012b, Tremblay2016} found multi-wavelength evidence of both large- and small-scale mechanical feedback in the BCG of galaxy cluster Abell\,2597. They reported an extensive kpc-scale X-ray cavity network, with multiple rising buoyant bubbles, the largest of which coincides in both linear extent and position angle with the radio jet \citep{Tremblay2012a}. \textit{Hubble Space Telescope (HST)} and \textit{Herschel} observations reveal ongoing star formation co-spatial with knots in the X-ray emission \citep{Tremblay2012b}. Atacama Large Millimetre/submillimetre Array (ALMA) $^{12}$CO(2--1) observations further add to this picture, exposing chaotic cold accretion onto the central SMBH (by revealing absorption features in the AGN continuum caused by clouds moving inwards towards the SMBH; \citealt{Tremblay2016}) and a molecular outflow associated with the jet itself.

Here we report on molecular gas observations of NGC\,0708, the BCG in the low-mass galaxy cluster Abell\,262, itself part of the Perseus-Pisces galaxy supercluster. NGC\,0708 lies $58.3\pm5.4$\,Mpc away (estimated using infrared surface brightness fluctuations; \citealt{Jensen2003}).  
It is a giant elliptical galaxy with a weak dust lane \citep{Ebneter1985,Wegner1996} and an effective radius of 33\arcsec\ ($\approx\,9.3$\,kpc; \citealt{Wegner2012}). See Fig. \ref{fig:708_XRayJet} for a \textit{HST} image of NGC\,0708.
Abell\,262 was identified as having an X-ray emitting ICM by \citet{Jones1984}, and \citet{Stewart1984} measured the cooling time to be $1.3\times10^{9}$\,yr, smaller than the age of the universe so that the cluster is expected to form a cooling flow. The 20-cm observations of \citet{Parma1986} revealed a double-lobed, 'S'-shaped jet and led to the classification of NGC\,0708 as a weak Fanaroff--Riley Class\,I radio source \citep{Blanton2004}. The top panel of Fig. \ref{fig:708_XRayJet} also has 330\,MHz continuum observations from \citet{Clarke2009} overlaid (blue contours), to show the shape and orientation of the large-scale jet.
Analysis of \textit{Chandra} observations revealed a hole or bubble within the ICM, co-spatial with the eastern lobe of the jet \citep{Blanton2004}. \citet{Clarke2009} found additional 3-6 kpc radius cavities at differing position angles within the X-ray gas, and at a range of radial distances from the BCG (8 -- 29 kpc), indicating multiple episodes of AGN activity from a precessing SMBH jet. They concluded that the total AGN emission should be capable of counteracting the cooling flow over several outbursts. 
Using their multi-frequency observations of NGC\,0708, \citet{Clarke2009} also calculated the radio spectral index ($\alpha$) from $235$ to $610$\,MHz, finding the spectrum to be flat in the core ($\alpha=-0.5$), typical of new particles in a jet. 
They also estimated a lower limit on average outburst repetition timescales in Abell\,262 to be $\tau_{\mathrm{rep}}\ge28$\,Myr.

NGC\,0708 was thus observed to have large-scale feedback affecting the hot gas. In this work, we show that the cold interstellar medium (ISM) is also being affected on small scales by the feedback cycle in this object, as expected if AGN feeding/feedback is a multiscale self-regulated process. In Section \ref{sec:obs}, we present new ALMA and enhanced Multi-Element Radio Linked Interferometer Network ({e-MERLIN}) observations of NGC\,0708. We present our analysis of these observations in Section \ref{sec:results}. In Section \ref{sec:discussion} we discuss our results and compare NGC\,0708 to other galaxies. We conclude in Section \ref{sec:conc}.

\begin{figure}
	\includegraphics[width=\columnwidth,trim=0cm 0cm 0cm 0cm,clip]{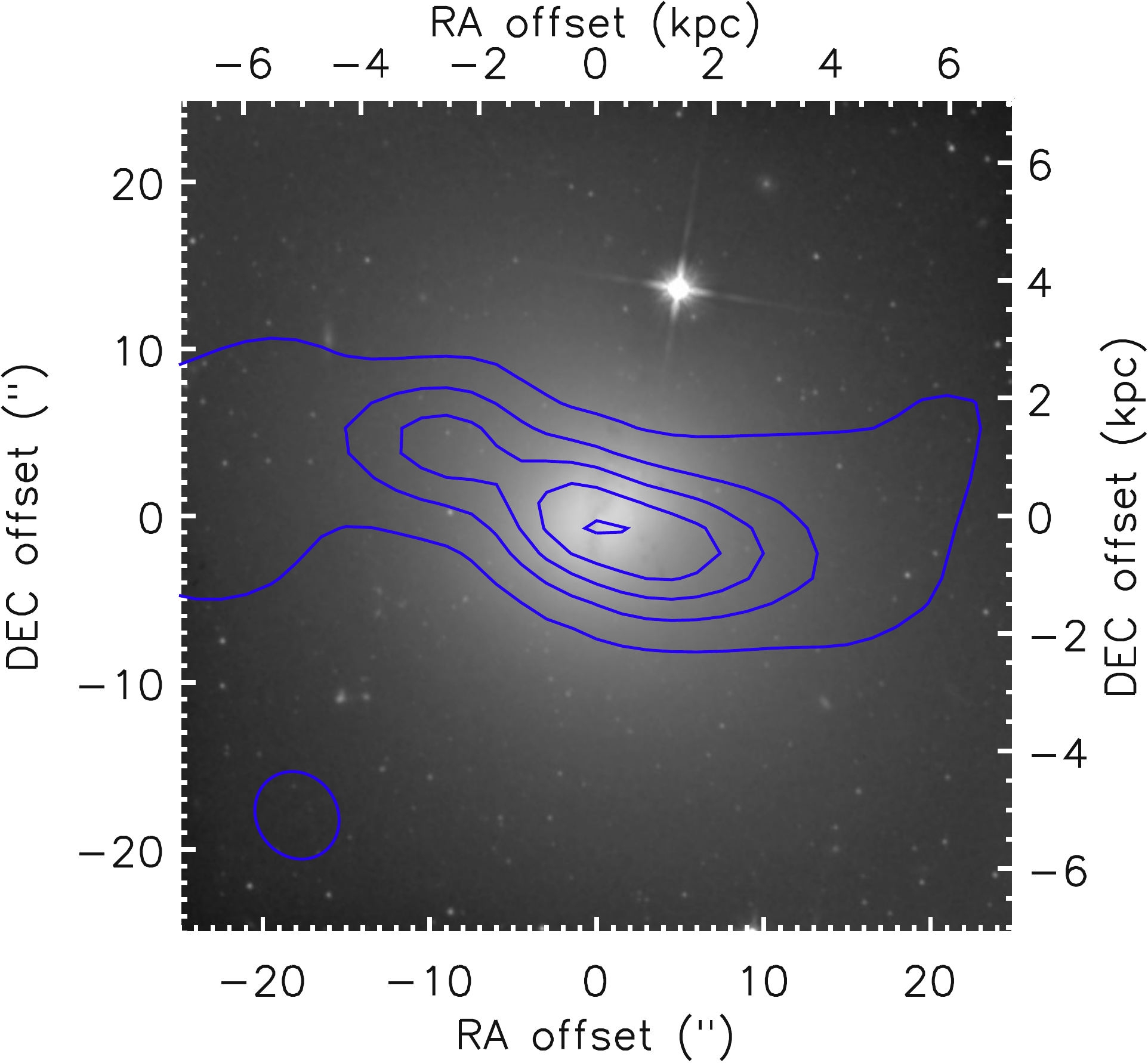}\vspace{0.1cm}
	\includegraphics[width=\columnwidth,trim=87.0cm 0cm 0.0cm 0cm,clip]{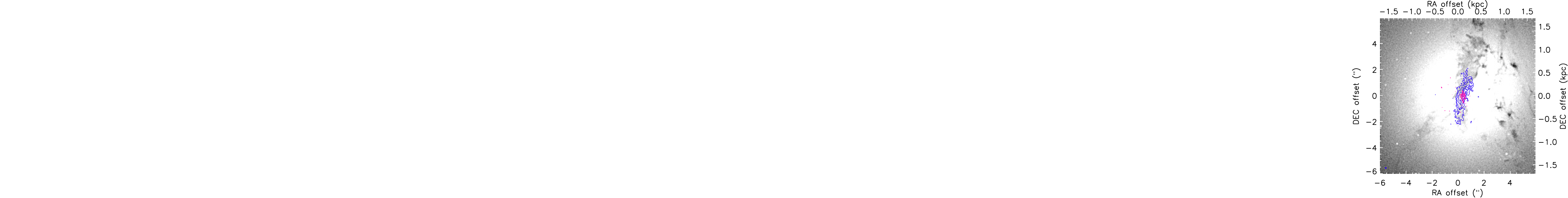}
	\caption{\textit{Top:} Large-scale (50\arcsec$\times$50\arcsec\ or $\approx14\times14$\,kpc$^{2}$) \textit{HST} Wide Field Camera 3 F110W image of NGC\,0708, with 330\,MHz continuum Very Large Array (VLA) contours from \protect  \cite{Clarke2009} overlaid in blue. \textit{Bottom}: Small-scale (12\arcsec$\times$12\arcsec\ or $3.2\times3.2$\,kpc$^{2}$) \textit{HST} combined Advanced Camera for Surveys and Wide Field Camera F435W image of NGC\,0708, with our CO(2--1) integrated intensity (0\farcs088 or $\approx$25\,pc resolution) contours overlaid in blue and 236\,GHz continuum contours overlaid in magenta. The synthesised beam of the radio/millimetre data is shown in the bottom-left corner of each image.}
	\label{fig:708_XRayJet}
\end{figure}

\section{Observations}
\label{sec:obs}

{NGC\,0708 has been observed many times in the CO wavebands using the Institut de Radioastronomie Millim\'{e}trique (IRAM) 30-m telescope, firstly by \citet{Edge2001} who published 3 observations, two at 113.45 GHz (with a 21\farcs2 beamsize) and one at 226.9\,GHz (with a 10\farcs6 beamsize). \citet{Edge2001} calculate a molecular gas mass of $(9\pm1.3)\times10^{8}$\msun\ and a beam temperature ratio of CO(1--0)/CO(2--1)=0.25.  \citet{Salome2003} only detected CO(1--0) and calculated a lower molecular gas mass of $(2\pm0.3)\times10^{8}$\msun\ due to identifying a line with a smaller width. Finally \citet{Ocana2010} also detected CO(1--0) and calculate a mass of $(4.5\pm1.1)\times10^{8}$\msun (where here we have converted these estimates to a common Milky Way-like CO-to-H$_{2}$ conversion factor of 2$\times$10$^{20}$ cm$^{-2}$ (K \kms)$^{-1}$)}.

{CO(2--1) in NGC\,0708 was observed three times with ALMA (using different array configurations) as part of the mm-Wave Interferometric Survey of Dark Object Masses (WISDOM) project, aiming to measure its central SMBH mass. Previous work (e.g. \citealt{Woo2002};  \citealt{Donato2004}) suggested an SMBH mass $M_{\mathrm{BH}}\approx2.9\times10^{8}$\msun, although this is highly uncertain. \citet{Olivares2019} published our intermediate-resolution observations from ALMA in a study of filaments in cool-core clusters. The observations at 0\farcs$95\times0$\farcs61 ($\approx270\times170$\,pc$^{2}$) show no filaments but instead a slightly warped rotating kiloparsec scale disc of molecular gas. Here we study this source in detail by including both higher and lower-resolution observations from WISDOM.}

\subsection{ALMA observations}
\label{sec:ALMA}
\begin{figure*}
	\includegraphics[width=\textwidth,trim=0cm 0cm 0cm 0cm,clip]{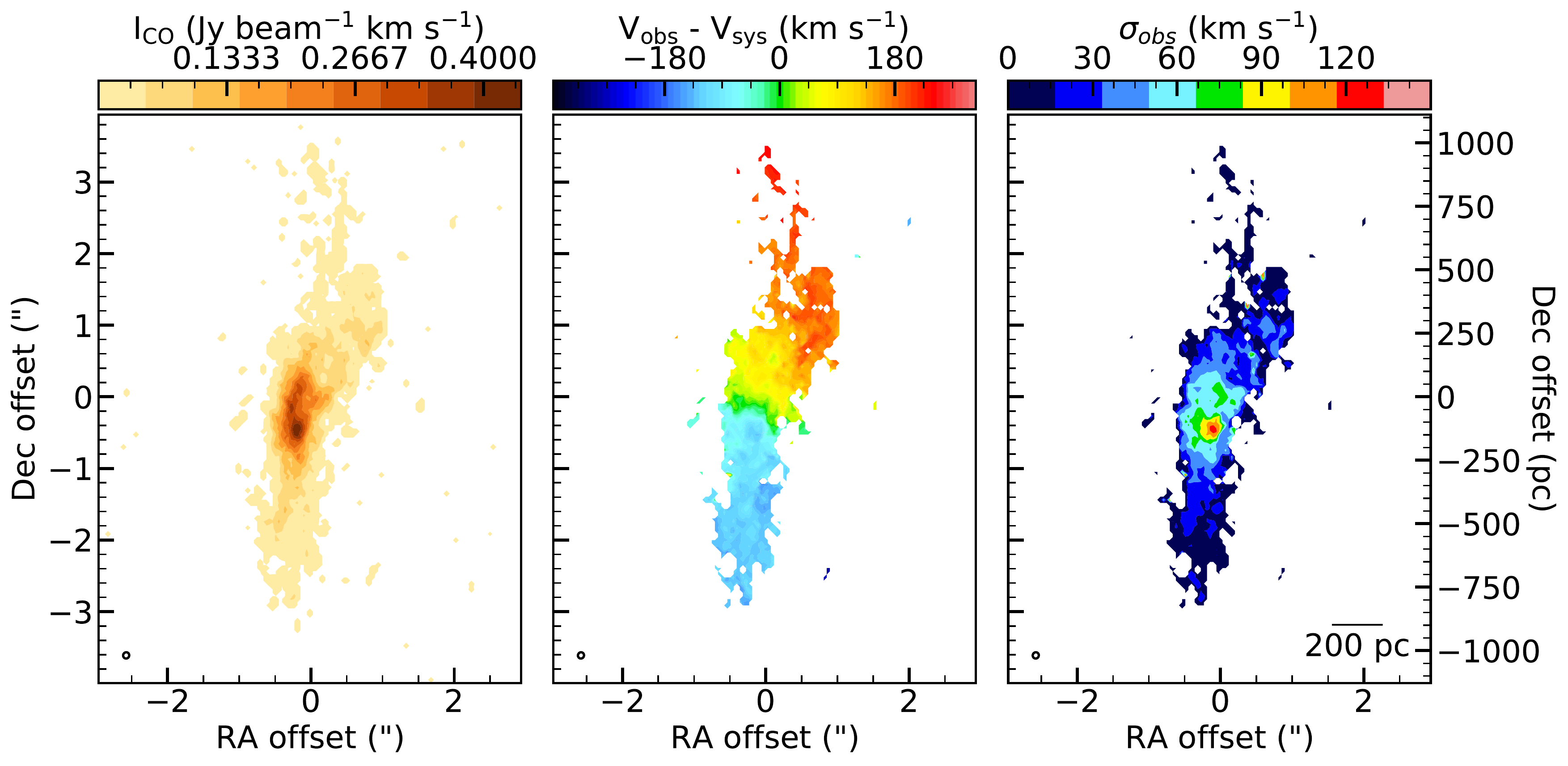}
    \caption{$^{12}$CO(2--1) moment maps of NGC\,0708. \textit{Left:} moment zero (integrated intensity) map.  \textit{Centre:} moment one (intensity-weighted mean line-of-sight velocity) map, assuming a systemic velocity $V_\mathrm{{sys}}=4750$\kms. \textit{Right:} moment two (observed intensity-weighted line-of-sight velocity dispersion) map. Note the off-centre velocity dispersion peak. 
    The ellipse in the bottom-left corner of each panel shows the synthesised beam (0\farcs$088\times0$\farcs083 or $\approx25\times23$\,pc$^{2}$). RA and Dec. offsets are relative to the central peak of the 236\,GHz continuum emission, located at ICRS position $\mathrm{RA}=01^{\mathrm{h}}52^{\mathrm{m}}46$\fs46 and $\mathrm{Dec.}=+36$\degr09\arcmin06\farcs46.}
    \label{fig:708_mom01} 
\end{figure*}

Using ALMA, we observed the $^{12}$CO(2--1) line in NGC\,0708, first under programme 2015.1.00598.S at moderate angular resolution (0\farcs52 or $\approx146$\,pc) on June 27th 2016 {(published in \citealt{Olivares2019})}, and then under programme 2017.1.00391.S at 0\farcs25 ($\approx70$\,pc) resolution on November 12th 2017 and 0\farcs03 ($\approx8.5$\,pc) resolution on September 19th 2018. The science target integration times for these were 11, 19 and 37\,min, respectively. The baselines ranged from 15\,m to 14\,km, yielding a maximum recoverable scale of 7\arcsec\ ($\approx2$\,kpc), sufficient to cover the extent of the majority of the dust features in this source. For all observations a 1870\,MHz ($\approx2500$\kms) correlator window was centred at 226.8\,GHz (the redshifted $^{12}$CO(2--1) line frequency) with a raw channel width of $976.5$\,kHz ($\approx1.87$\kms).  
To detect continuum emission, three additional low-spectral resolution correlator windows were included, each with a bandwidth of $2$\,GHz.

The raw data were calibrated using the standard ALMA pipeline, as provided by the European ALMA Regional Centre staff. The calibrators used for all observations were J0237+2848 for flux and bandpass calibration and J0205+3212 for phase calibration. 
The three observation tracks were combined and imaged using the \textsc{Common Astronomy Software Applications} (\textsc{casa}; \citealt{McMullin2007}).
Continuum emission was detected, measured over the full line-free bandwidth, and then subtracted from the data in the $uv$--plane using the \textsc{casa} task \textsc{uvcontsub}. 
Both the line and continuum cube were cleaned and imaged using the \textsc{casa} task \textsc{tclean} and Briggs weighting with a robust parameter of 0.5. Both were then primary-beam corrected. 
The imaging achieved a synthesised beam size of full-width at half-maximum (FWHM) 0\farcs$088\times0$\farcs083 ($\approx25\times23$\,pc$^{2}$) for the $^{12}$CO(2--1) line and 0\farcs$088\times0$\farcs087 ($\approx25\times25$\,pc$^{2}$) for the continuum. 
To produce the final three-dimensional CO(2--1) RA-Dec.-velocity data cube, the data were binned to 10\kms\ channels and 0\farcs$035\times0$\farcs035 spaxels ($\approx3$\,spaxels across the synthesised beam major axis ensures Nyquist sampling). 
This CO(2--1) cube has a root mean square (RMS) noise of 0.41\,mJy\,beam$^{-1}$ in each (emission free) 10\kms\ channel. The final continuum image has an RMS noise of 16$\mu$Jy beam$^{-1}$.

\subsection{e-MERLIN observations}
\label{sec:eM}
We also observed NGC\,0708 with {e-MERLIN}, to identify if any small-scale radio structures are present (e.g. a restarted radio jet from the recurrent, precessing AGN in the core of this galaxy). 
NGC\,0708 was thus observed twice with {e-MERLIN}, the data providing sensitivity to 5\,GHz emission distributed on the same angular scales as our 236\,GHz data. 
The {e-MERLIN} data were processed through the standard {e-MERLIN} \textsc{casa} pipeline (eMCP\footnote{\url{https://github.com/e-MERLIN/eMERLIN_CASA_pipeline}}) by the {e-MERLIN} facility staff. The calibrators used were 0152+3616 for phase, 0319+4130 for pointing, 1331+3030 for flux and 1407+2827 for bandpass calibration. 
The total on-source integration time was 14.5\,hours.	

We additionally performed self-calibration to increase the sensitivity. The self-calibration involved 2 cycles, the first considering phase only, averaging over 240\,s intervals, the second with both phase and amplitude. 
We imaged the data in \textsc{casa} using the \textsc{tclean} task, with a Briggs weighting robust parameter of 0.5 to balance sensitivity and resolution. This yielded a synthesised beam size of 0\farcs$07\times0$\farcs03 ($\approx20\times8$\,pc$^{2}$) and a RMS noise of $0.12$\,mJy\,beam$^{-1}$.

\section{Results}
\label{sec:results}

\subsection{ALMA line emission}
\label{sec:LineEm}
The moment maps, shown in Fig. \ref{fig:708_mom01}, were created using the smooth-mask technique \citep[e.g.][]{Dame2011}. The mask was generated by taking a copy of the cleaned, primary beam-corrected cube and smoothing it, first spatially using a Gaussian of FWHM equal to that of the synthesised beam, and then spectrally using a Gaussian of FWHM of 4 channels. We then select pixels with an amplitude in the smoothed cube greater than 5 times the RMS noise in that cube. The mask was then applied to the un-smoothed cube to create the moment maps. Having said that, all quantitative analyses reported in this paper were performed using the un-smoothed, un-masked cube. The moment maps are shown only for illustrative purposes.

The zeroth (integrated intensity) and first (intensity-weighted mean line-of-sight velocity) moment maps reveal a rotating but warped molecular gas disc (see left and central panels Fig. \ref{fig:708_mom01}). The moment zero is also shown in the bottom panel of Fig. \ref{fig:708_XRayJet} (blue contours), overplotted on a \textit{HST} image, revealing that the molecular gas is coincident with dust features. 
The second moment (intensity weighted line-of-sight velocity dispersion; right panel Fig. \ref{fig:708_mom01}) shows evidence of disturbance, with an off-centre peak significantly away ($\approx0$\farcs4 or $\approx113$\,pc) from the AGN position (that can also be independently measured from our data using the 236\,GHz continuum emission; see Section \ref{sec:cocont}). 
A major-axis position-velocity diagram (PVD; Fig. \ref{fig:708_PVD}) was created by taking a 5-pixel wide pseudo-slit across the kinematic major axis of the cube, at a position angle of 349\degr. This position angle was estimated by eye and agrees with that found by \citet{Pandya2017} for the ionised-gas disc traced by [\ion{O}{iii}].  
On the approaching side, the PVD has a sharp increase in velocity at a distance of $\approx0$\farcs4 ($\approx113$\,pc) from the centre, co-spatial with the aforementioned increase in velocity dispersion (see the right panel of Fig. \ref{fig:708_mom01} and the blue ellipse in Fig. \ref{fig:708_PVD}). The properties of this blue-shifted emission will be considered further in Section \ref{sec:bluecomp}.

The global spectrum, shown in Fig. \ref{fig:708_spec}, was created by binning the data to 20\kms\ channels and then integrating over the whole molecular gas disc, i.e. a 6\arcsec$\times$6\arcsec\ ($1.7\times1.7$\,kpc$^{2}$) area of the cube. This spectrum clearly shows the characteristic double-horned profile of a rotating disc, but with hints of an additional blue-shifted wing (highlighted by the magenta ellipse). 
The total $^{12}$CO(2--1) flux detected in NGC\,0708 is 57.9$\pm$0.1(stat)$\pm$5.8(sys) Jy\kms. The uncertainties quoted are the $1\sigma$ statistical uncertainty, and the $\approx$10~percent systematic flux calibration uncertainty, that typically dominates over the statistical uncertainty. 

We estimate the H$_2$ gas mass present in this system, using the following standard equation:
\begin{equation}
M_\mathrm{H_2} = 2m_{\mathrm{H}} \frac{\lambda^2}{2 k_\mathrm{b}}\, X_{\rm CO}\, D_{\rm L}^2\, R_{\rm 2-1} \, \int{S_v \delta \mathrm{V}},
\end{equation}
\noindent where $m_\mathrm{H}$ is the mass of a hydrogen atom, $\lambda$ is the rest wavelength of the observed molecular transition, $k_\mathrm{b}$ is Boltzmann's constant, $D_\mathrm{L}$ is the luminosity distance, $R_{\rm 2-1}$ = $T_{\mathrm{b, CO(1-0)}}/T_{\mathrm{b, ref}}$ is the line ratio (measured in beam temperature units) between the reference CO transition observed and the ground state CO(1-0) line, $\int{S_v \delta \mathrm{V}}$ is the integrated CO flux density and  X$_{\rm CO}$ is your CO-to-H$_2$ conversion factor of choice. This can be simplified to:
\begin{eqnarray}
\left(\frac{M_\mathrm{H_2}}{M_{\odot}}\right) = 7847\,J_{\rm upper}^{-2}\,X_{\rm CO, 2\times10^{20}}\,R_{\rm 2-1}  \left(\frac{D_{\rm L}}{\mathrm{Mpc}}\right)^{2} \left(\frac{ \int{S_v \delta \mathrm{V}}}{\mathrm{Jy\,km\,s}^{-1}}\right),\\
\mathrm{where\,\,\,} X_{\rm CO, 2\times10^{20}} = \frac{X_{\rm CO}}{2\times10^{20} \mathrm{cm}^{-2} (\mathrm{K\,km\,s}^{-1})^{-1}}, \nonumber
\label{co2h2}
\end{eqnarray}
\noindent and $J_{\rm upper}$ is the upper state rotational quantum number for the transition observed (here $J_{\rm upper}=2$). The other symbols are as defined above. 

We here assume a typical Milky Way-like CO-to-H$_{2}$ conversion factor of 2$\times$10$^{20}$ cm$^{-2}$ (K \kms)$^{-1}$ \citep{Dickman:1986jz} (equivalent to $\alpha_{\mathrm{CO}}\approx4.6$\msun\,(K\kms)$^{-1}$\,pc$^{-2}$) and that the gas is has a line ratio {$T_{\mathrm{b, CO(2-1)}}/T_{\mathrm{b, CO(1-0)}}=0.25$ (\citealt{Edge2001})}, and thus $R_{2-1}=4$. The total flux therefore corresponds to a total molecular gas mass {$M_{\mathrm{tot}}=(1.5\pm0.01$(stat)$\pm$0.26(sys))$\times10^{9}$\msun}, which is slightly higher than that found by previous single dish measurements (\citealt{Edge2001}; \citealt{Salome2003}; \citealt{Ocana2010}) suggesting that our combined ALMA observations do not resolve out a significant fraction of the extended flux in this object, despite high angular resolution. We note that the molecular gas mass derived here is different from that derived from a subset of these same data in \cite{Olivares2019}. This difference arises because of different assumptions about the $T_{\mathrm{b, CO(2-1)}}/T_{\mathrm{b, CO(1-0)}}$ ratio and $X_{\rm CO}$.

\begin{figure}
	\includegraphics[width=\columnwidth]{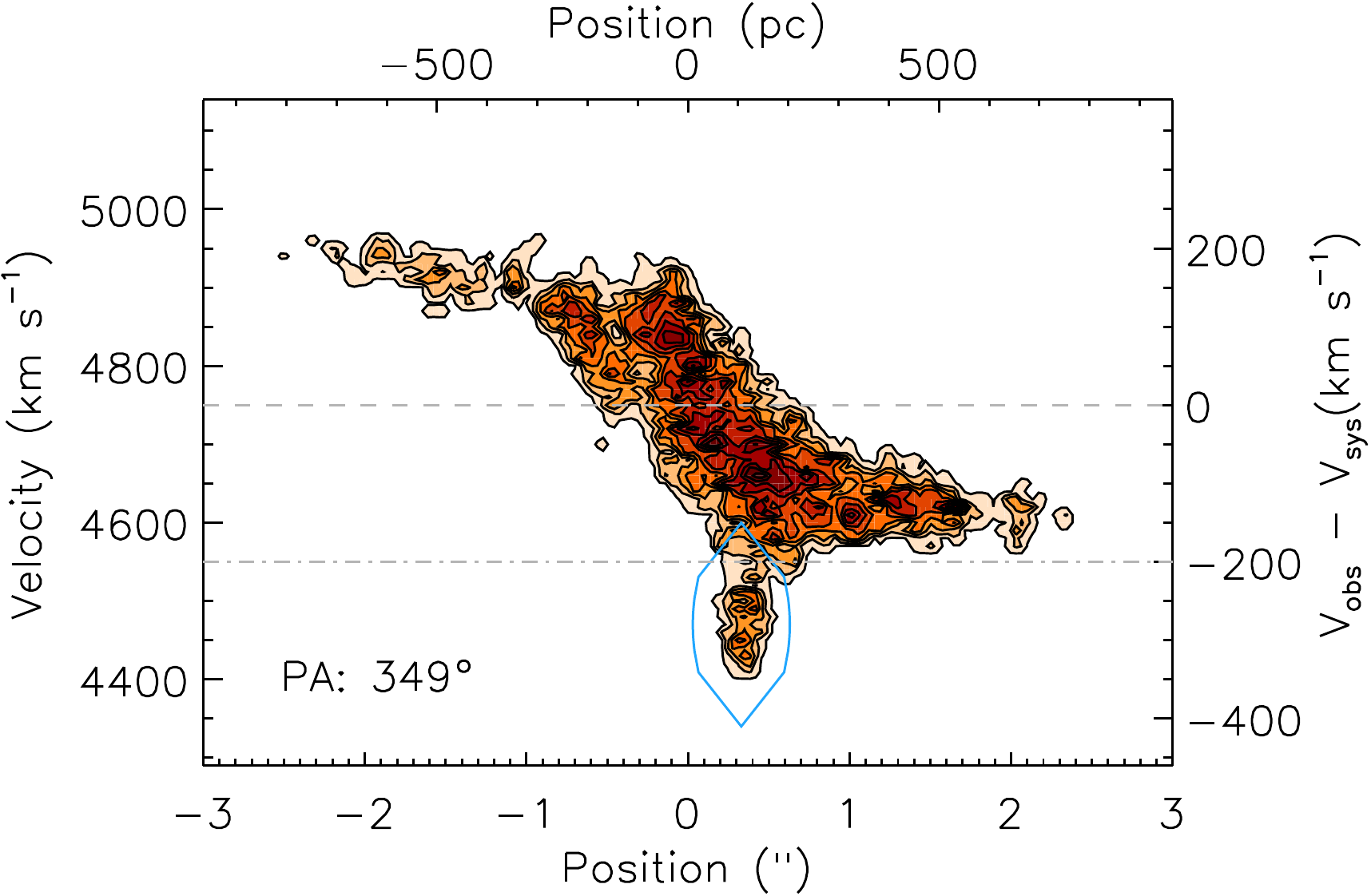}
	\caption{$^{12}$CO(2--1) kinematic major-axis position-velocity digram of NGC\,0708, taken at a position angle of 349\degr. The grey dashed line denotes the systemic velocity, $V_{\mathrm{sys}}=4750$\kms. The grey dot-dashed line denotes the velocity cut used to isolate the blue-shifted feature ($V_{\mathrm{obs}}=4550$\kms; see Section \ref{sec:bluecomp}), itself indicated by the cyan polygon.}
	\label{fig:708_PVD}
\end{figure}
\begin{figure}
	\includegraphics[width=\columnwidth]{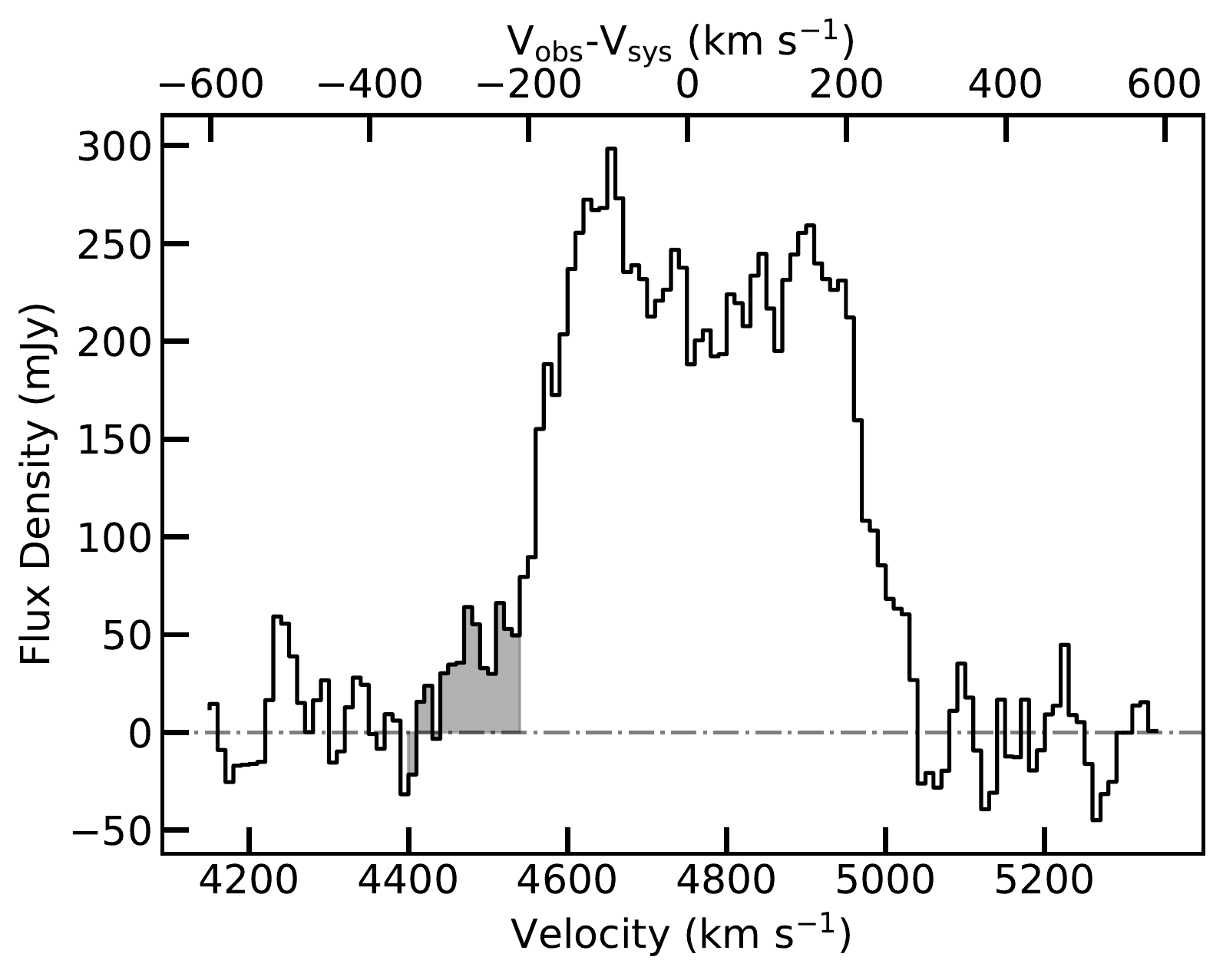}
	\caption{$^{12}$CO(2--1) integrated spectrum of NGC\,0708 smoothed to an effective resolution of 20\kms, showing the characteristic double-horned shaped of a rotating disc. The anomalous blue-shifted wing is indicated by the shaded grey region.}
	\label{fig:708_spec}
\end{figure}

\subsection{ALMA continuum emission}

\label{sec:cocont}
As mentioned in Section \ref{sec:ALMA}, NGC\,0708 also has 1\,mm continuum emission, detected by ALMA at a mean frequency of 236\,GHz in the three low-resolution spectral windows and the line-free channels of the high-resolution spectral window. The emission is resolved and has an extension to the South, clearly revealed in the left-hand panel of Fig. \ref{fig:CM2spec}, showing the 236\,GHz continuum emission (magenta contours) overlaid on the CO(2--1) velocity dispersion map.
This highlights both the approximate coincidence between the extension of the continuum emission and the peak of the velocity dispersion, and the offset of that peak from the AGN position (i.e. the centre of the continuum emission). The magenta arrows in the left-hand panel of Fig. \ref{fig:CM2spec} show the direction of the large-scale 330\,MHz jet, to highlight the difference of orientation between that and the 236\,GHz emission.
The total 236\,GHz continuum flux is $32.3\pm0.2$\,mJy. Excluding the nucleus (by subtracting a point-source fit to the central peak from the total flux), the extended continuum emission component has a flux of 1.0$\pm$0.2\,mJy (where both figures are quoted with $1\sigma$\ statistical uncertainties).

 \begin{table*}
\caption{Model parameters with their priors, best-fitting values and statistical uncertainties.}
\begin{center}
\begin{tabular*}{0.75\textwidth}{@{\extracolsep{\fill}}l r c r r}
\hline
Parameter & \multicolumn{3}{c}{Prior} & Best fit with 1$\sigma$ error (68\%) \\
(1) &  \multicolumn{3}{c}{(2)} & (3) \\
\hline
Galaxy parameters:&&&& \\\hline
Modelled region flux (Jy km s$^{-1}$) &   10& $\xrightarrow[]{\rm uniform}$ &  50 &   23.1$\pm$1.7    \\
Asymptotic velocity ($V_{\rm max}$; km s$^{-1}$) &  100& $\xrightarrow[]{\rm uniform}$ & 200 &    160 $\pm$21\\
Velocity turnover radius  ($R_{\rm turn}$; arcsec) &  0.01& $\xrightarrow[]{\rm uniform}$ & 1.0 &    0.21 $\pm$0.11\\
Surface brightness scale radius  ($R_{\rm turn}$; arcsec) &  0.1& $\xrightarrow[]{\rm uniform}$ & 1.0 &    0.49 $^{+0.07} _{-0.05}$\\
Scale height ($z_{\rm scale}$; pc) &  0.0& $\xrightarrow[]{\rm uniform}$ & 275 &    71 $\pm$16 \\
Gas velocity dispersion ($\sigma$; \kms) &   1& $\xrightarrow[]{\rm uniform}$ &100 &   66 $\pm$5  \\
\\
Nuisance parameters: &&&&\ \\\hline
Position angle  ($^\circ$) & 330& $\xrightarrow[]{\rm uniform}$ & 360  &   347 $\pm$3  \\
Inclination  ($^\circ$) &  70& $\xrightarrow[]{\rm uniform}$ & 89 &   76.6 $\pm$3.3  \\
Centre X offset (arcsec) &  $-$1.0& $\xrightarrow[]{\rm uniform}$ &  1.0 &   0.04 $\pm$0.02 \\
Centre Y offset (arcsec) &  $-$1.0& $\xrightarrow[]{\rm uniform}$ &  1.0 &     -0.09 $\pm$0.03 \\
Centre velocity offset (\kms) & $-$20.0& $\xrightarrow[]{\rm uniform}$ & 20.0 &   4.47 $\pm$6.31\\ \hline
\end{tabular*}
\parbox[t]{0.75\textwidth}{ \textit{Notes:} Column 1 lists the fitted model parameters, while Column 2 lists the prior for each. All priors are uniform in linear space between the limits given. The X, Y and velocity offset nuisance parameters are defined relative to the International Celestial Reference System (ICRS) position $\mathrm{RA}=01^{\mathrm{h}}52^{\mathrm{m}}46$\fs46 and $\mathrm{Dec.}=+36$\degr09\arcmin06\farcs46, and the systemic (barycentric) velocity of the galaxy $V$=4750 km s$^{-1}$.}
\end{center}
\label{fittable}
\end{table*}

\subsection{e-MERLIN 5\,GHz continuum emission}
\label{sec:eMcont}
The extension of the 236\,GHz continuum emission in NGC\,0708 is perpendicular to the large-scale jet (as traced by 330\,MHz emission; see Figs. \ref{fig:708_XRayJet} and \ref{fig:CM2spec}) and this prompted us to obtain additional 5\,GHz continuum data, to ascertain if the 236\,GHz continuum emission is from a small (potentially restarted) jet. These observations are described in Section \ref{sec:eM}.

We detect a 5\,GHz point source at the expected position of the SMBH in NGC\,0708. We confirmed this source is spatially unresolved using the \textsc{casa} task \textsc{imfit}, that fits a Gaussian to the image, deconvolved from the synthesised beam. 
The integrated flux at 5\,GHz is $5.25\pm0.21$\,mJy ($1\sigma$ statistical uncertainty). This is $\approx5$ times smaller than that measured by \citet{Clarke2009} at 5\,GHz with the VLA on $\approx4$\arcsec\ ($\approx1.1$\,kpc) scales and $6-8$ times smaller than single-dish 5\,GHz measurements ($\approx2$\farcm6 or $\approx44$\,kpc scales, \citealt{Andernach1980}; $\approx10$\arcmin\ or $\approx170$\,kpc scales, \citealt{Gregory1996}). The disparity between these measurements and ours suggests that significant 5\,GHz emission is associated with the large-scale radio jet, that we resolve out here. There is no obvious small-scale (restarted) jet visible at 5\,GHz, although our observations are only deep enough to detect components with a 5\,--\,236\,GHz spectral index $<-0.36$ (at 3\,$\sigma$). If a young jet is present in this source this suggests that its 5\,--\,236\,GHz spectral index is shallow or inverted.

\subsection{Blue-shifted component}
\label{sec:bluecomp}

Our ALMA data enable us to spatially and kinematically separate distinct components of the molecular gas distribution. The velocity dispersion map (right panel of Fig. \ref{fig:708_mom01}) and PVD (Fig. \ref{fig:708_PVD}) clearly indicate two kinematically-distinct components are present in NGC\,0708, a disturbed but regularly rotating disc and a blue-shifted `spike' feature. Here we investigate this anomalous blue-shifted feature.

\subsubsection{Kinematic model}
\label{section:kinmsdiscmodel}

We begin by kinematically modelling the disc component at the centre of NGC\,0708, to reveal the full extent of the anomalous emission. To construct this model we utilise the \textsc{Python} implementation of the \textsc{KinMS}\footnote{\url{https://github.com/TimothyADavis/KinMSpy}} kinematic modelling code \citep{2013MNRAS.429..534D,2020ascl.soft06003D}, and fit the model to our data using the Bayesian Markov-Chain Monte Carlo (MCMC) code \textsc{GAStimator}\footnote{\url{https://github.com/TimothyADavis/GAStimator}}.

We assume the mass distribution is axisymmetric (at least in the inner regions of interest here), and model the rotation curve of this system as a function of radius ($R$) with an arctangent function
\begin{equation}
V_{\rm rot}(R) = \frac{2 V_{\rm max}}{\pi} \arctan\left(\frac{R}{R_{\rm turn}}\right),
\label{v_as_func_r}
\end{equation}
where $V_{\rm max}$ is the asymptotic velocity, and $R_{\rm turn}$ is the turnover radius of the rotation curve. We further assume that the molecular disc has an exponential surface brightness profile both as a function of radius and vertical extension ($z$),
 \begin{equation}
I(R,z) = I_0 \exp^{\left(\frac{-R}{R_{\rm scale}}\right)}\exp^{\left(\frac{-|z|}{z_{\rm scale}}\right)},
\end{equation}
where $I_0$ is the central surface brightness, and $R_{\rm scale}$ is the exponential scale length and $z_{\rm scale}$ is the vertical scale height of the disc. These functional forms are very simple, and cannot reflect the full suite of rotation curves/surface brightness profiles present in nature. Despite this, here they allow us to construct a simple model which matches the data reasonably well. 

\begin{figure*}
	\includegraphics[height=6.2cm,trim=0cm 0cm 0cm 0cm,clip]{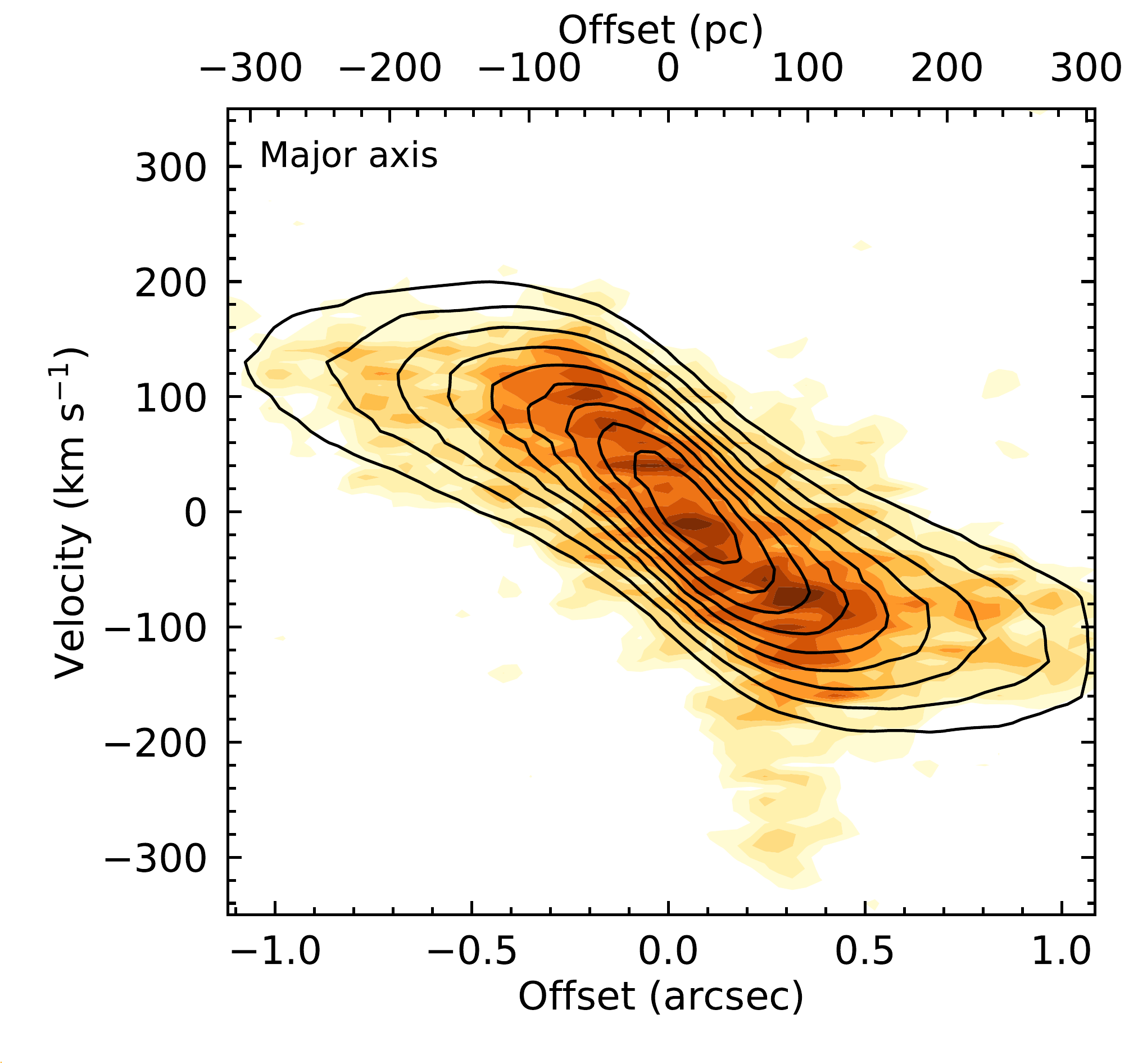}
		\includegraphics[height=6.2cm,trim=3.75cm 0cm 0cm 0cm,clip]{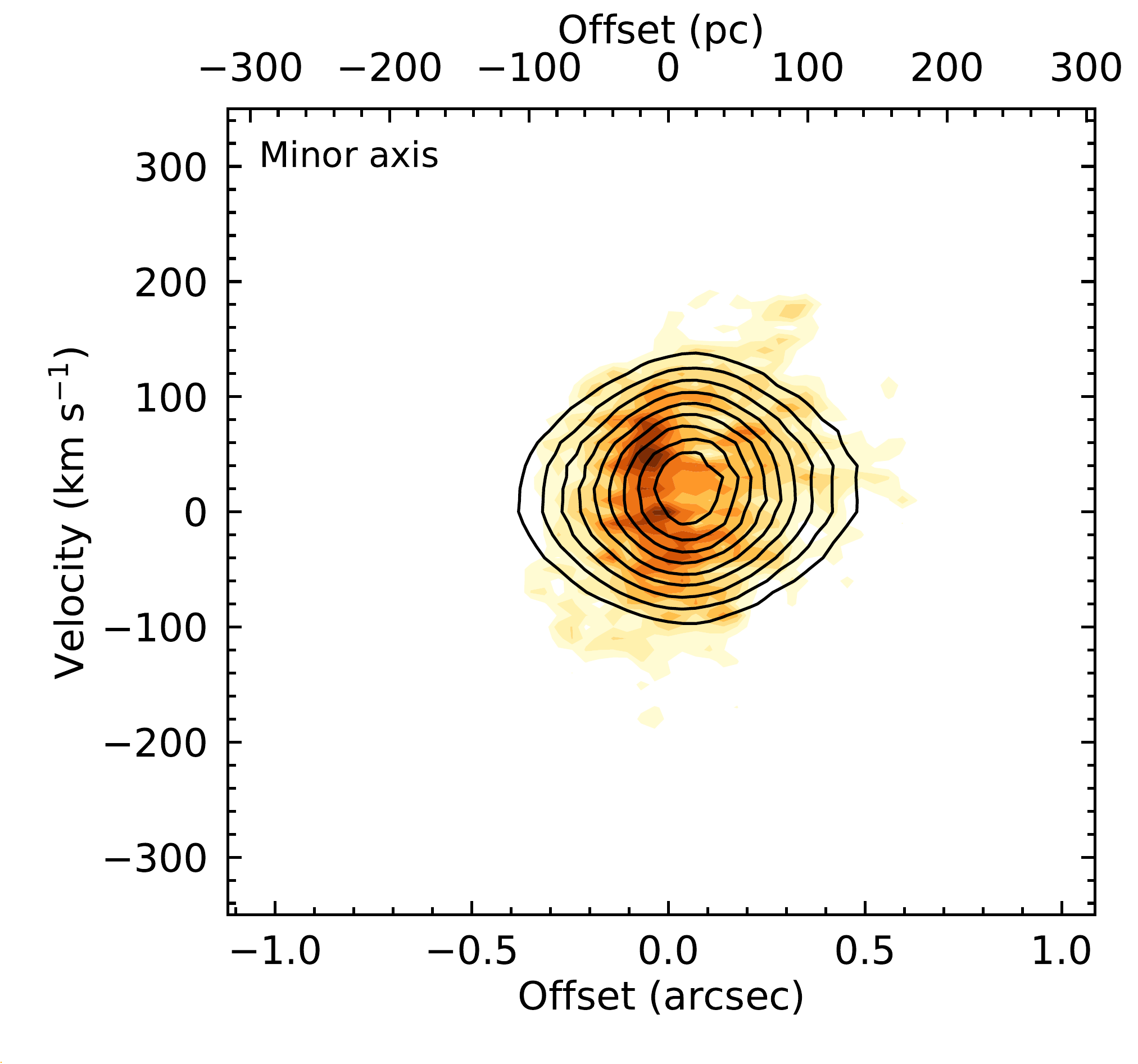}
		\includegraphics[height=6.2cm,trim=3.75cm 0cm 0cm 0cm,clip]{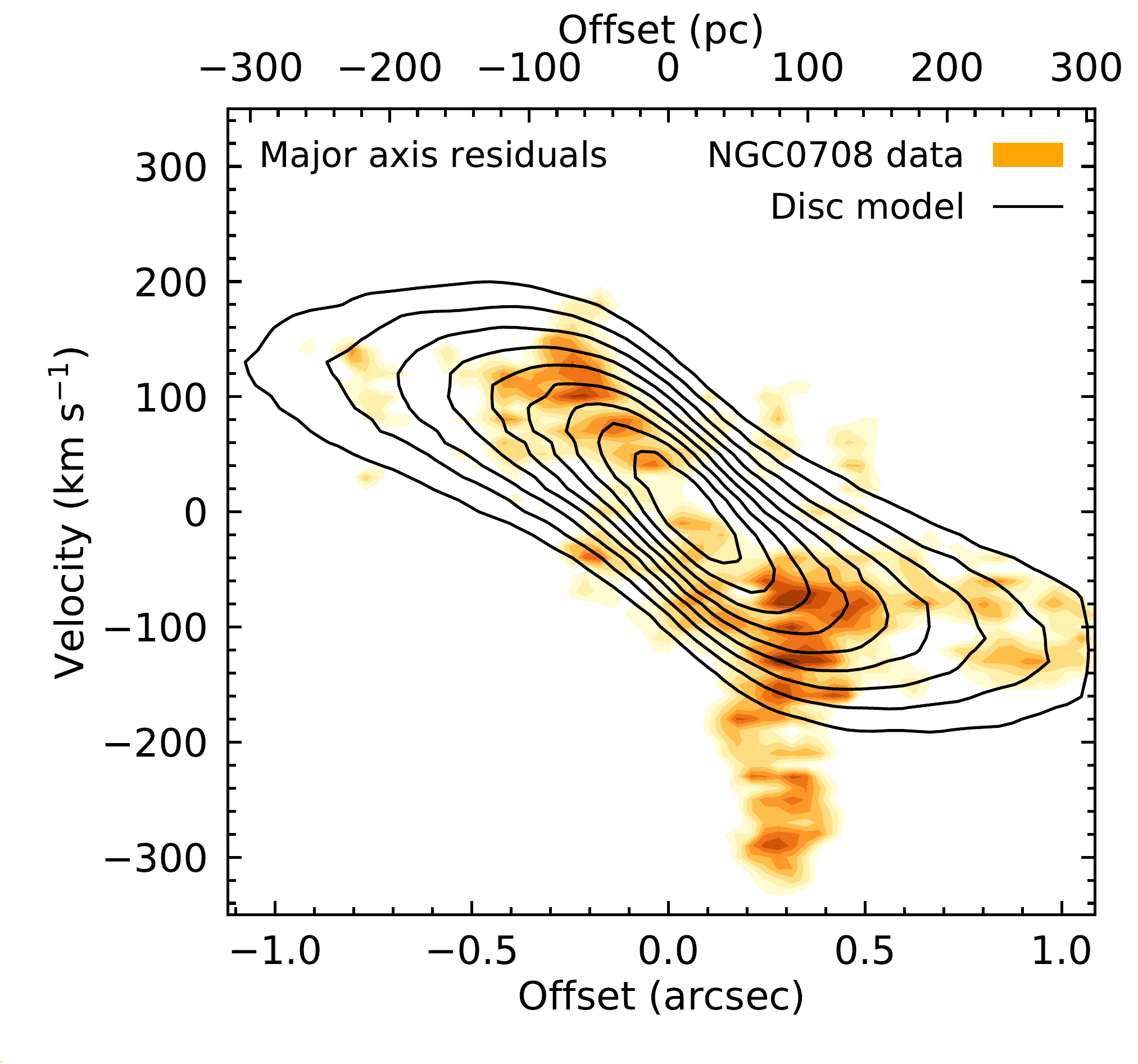}
	\caption{\textit{Left and Centre:} PVDs extracted from the NGC\,0708 CO datacubes along the kinematic major (349\degr; left column) and minor axis (259\degr; central column) of the system. The observed CO data is shown as orange filled contours, while the best-fitting \textsc{KinMS} model of a rotating disc is overlaid in black contours. \textit{Right:} Major axis PVD created from the residual (Data-model) cube (orange filled contours). Only positive residuals are shown due to the masking procedure used, but these are much larger than the negative residuals. The same disc model as present in the left panel is reproduced here, to guide the eye. This panel shows the full extent of the anomalous non-rotating  emission in this system.} 
	\label{fig:kinmsmodel}
\end{figure*}

\begin{figure*}
	\begin{center}
		\begin{tikzpicture}
		\node[anchor=south west,inner sep=0] (image) at (0,0) {
			\begin{minipage}[b]{1.05\columnwidth}	
			\begin{tikzpicture}
			\node[anchor=south west,inner sep=0] (image) at (0,0) {\includegraphics[height=7cm,trim=0cm 0cm 0cm 0cm,clip]{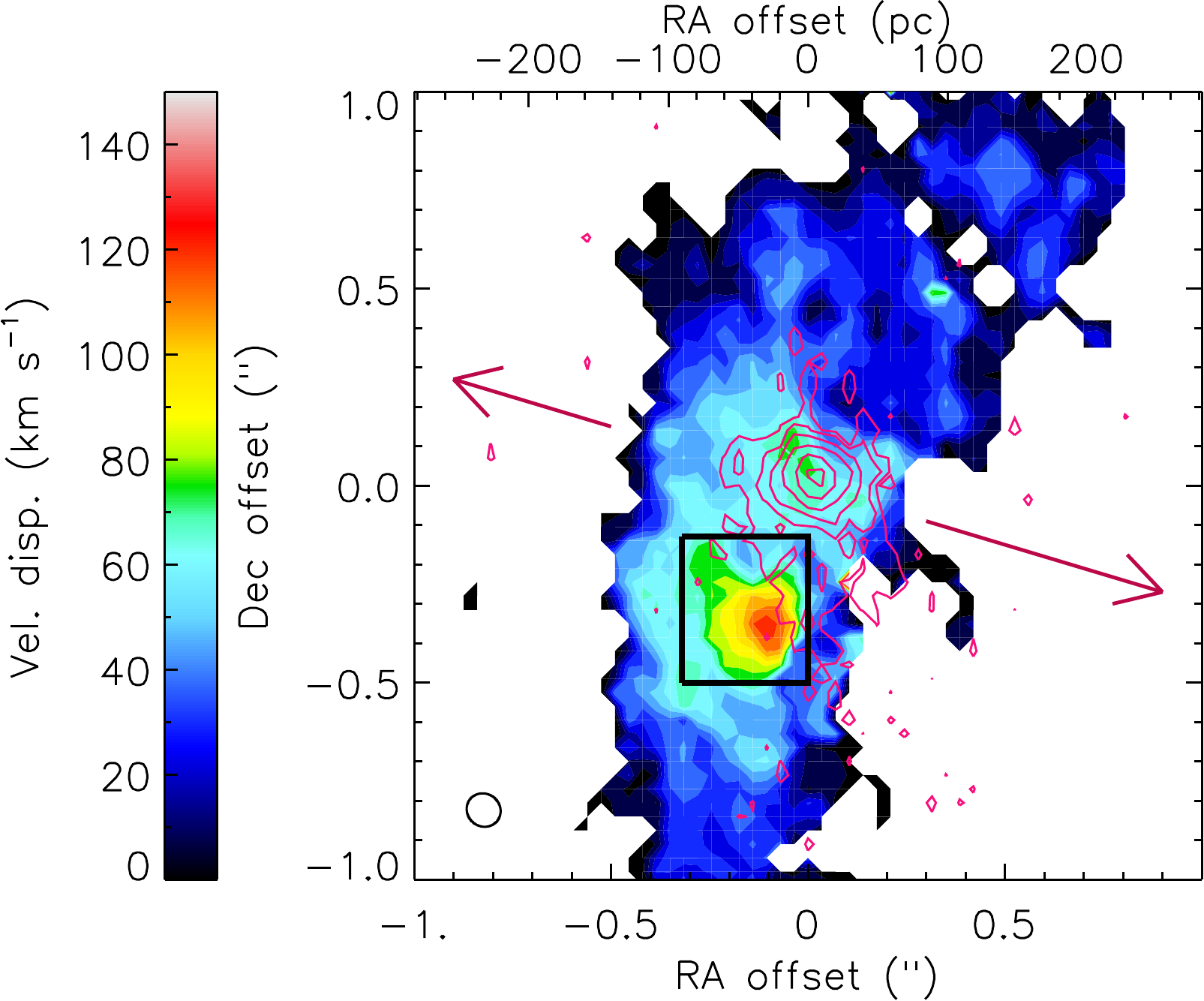}};
			\end{tikzpicture}
		\end{minipage}
		\begin{minipage}[b]{0.95\columnwidth}
			\begin{tikzpicture}
			\node[anchor=south west,inner sep=0] (image) at (0,0) {\includegraphics[height=7.5cm]{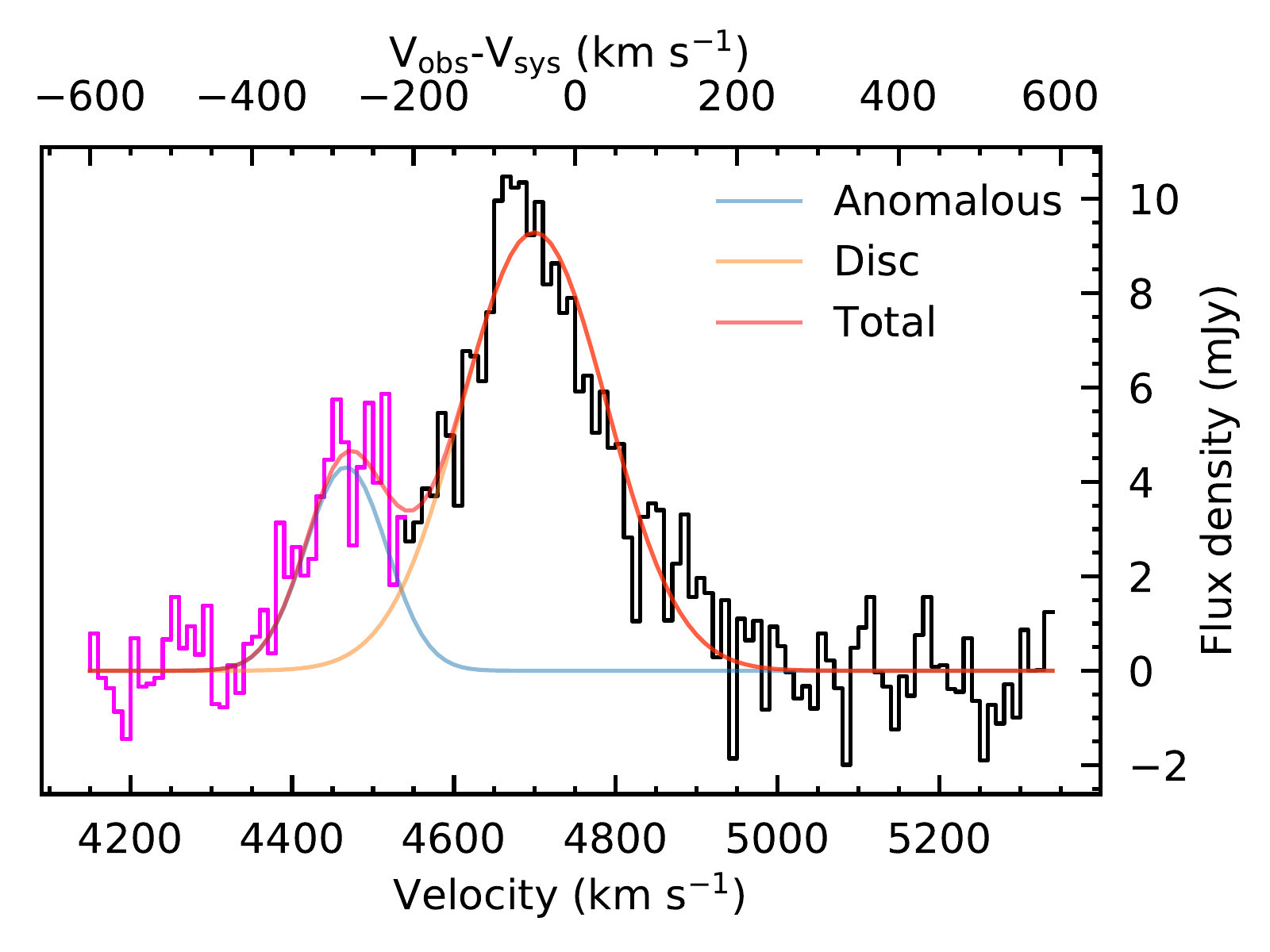}};
			\end{tikzpicture}
		\end{minipage}
	};
		\begin{scope}[x={(image.south east)},y={(image.north west)}]
		\draw[black, thick] (0.335,0.430) -- (0.547,0.845);	
		\draw[black, thick] (0.335,0.295) -- (0.547,0.165);
		\end{scope}
\end{tikzpicture}
\caption{{\textit{Left:} $^{12}$CO(2--1) moment 2 (intensity-weighted observed line-of-sight velocity dispersion) map of NGC\,0708, with 236\,GHz continuum emission isophotes overlaid in magenta. The magenta arrows indicate the direction of the large-scale jet traced by 330\,MHz emission. The extension of the 236\,GHz continuum emission matches well with the position of the velocity dispersion peak. \textit{Right:} $^{12}$CO(2--1) spectrum integrated over the spatial extent of the off-centred velocity dispersion peak (indicated by the black box in the left panel), with gas at anomalous velocities ($V_{\mathrm{obs}}<4550$\kms) indicated in magenta. Gaussian fits are shown in blue for the anomalous component, orange for the disc emission, and red for the sum of the two components. The anomalous emission is significantly blue-shifted from the galaxy systemic.}}
\label{fig:CM2spec}
\end{center}
\end{figure*}

\textsc{KinMS} takes these input functional forms, and allows us to construct a model datacube assuming the gas is in circular rotation around some unknown kinematic centre. We assume the gas has an intrinsic velocity dispersion ($\sigma$) which is another free parameter in our fitting process. One can only estimate the gas dispersion from full kinematic modelling, because the observed moment two map is often dominated by beam smearing, especially in the central regions where the rotation curve rises quickly. We further assume here that all the gas is distributed in a single plane (i.e. the disc has the same kinematic position angle and inclination to our light-of-sight throughout the disc). While these assumptions are unlikely to be valid in the more disturbed outer disc of this galaxy, they do produce a reasonable fit in the central regions (where dynamical times are very short).  We note that if we nevertheless allow radial variations of the inclination and position angle our best fit disc is consistent with being flat, suggesting that although it is morphologically lopsided, it may not be kinematically warped. Non-circular motions are almost certainly present within this gas disc, but here we aim to determine how much of the gas motion can be explained purely by regular rotation. 

As mentioned above, we use a Bayesian MCMC analysis technique to identify the model which best matches our data, and simultaneously estimate uncertainties. Indeed, this allows us to obtain samples drawn from the posterior distribution of each model parameter.
Our model, as described above, has a total of eleven free parameters (six key-parameters, and five nuisance parameters). The key parameters are the total flux of the system (which sets $I_0$ as described above), the two free parameters of the rotation curve, the scale radius of the molecular disc, its thickness in the vertical direction, and its velocity dispersion. The nuisance parameters the  position angle and inclination of the disc, and its kinematic centre (in RA, Dec, and velocity),  Each parameter has a prior, that we typically set as a box-car over a reasonable parameter range (an assumption of maximal ignorance). Details of all the priors are listed in Table \ref{fittable}.

Our MCMC procedure generates a model datacubes and compares them to our observed data using a simple log-likelihood ($\mathcal{L}$):
\begin{equation}
\mathcal{L}\propto\frac{-\chi^2}{2}, 
\end{equation}
where $\chi^2$ is the standard $\chi^2$ statistic.

As discussed in detail in \cite{2019MNRAS.485.4359S}, because our ALMA data are noisy, the $\chi^2$ statistic has an additional uncertainty associated with it, following the chi-squared distribution \citep{2010arXiv1009.2755A}. 
Systematic effects can produce variations of $\chi^2$ of the order of its variance \citep{2009MNRAS.398.1117V}, and ignoring this effect yields unrealistically small uncertainties. To mitigate this effect \cite{2009MNRAS.398.1117V} proposed to increase the $1\sigma$ confidence interval to $\Delta\chi^2=\sqrt{2N}$, where $N$ is the number of constraints. To achieve the same effect within our Bayesian MCMC approach we need to scale the log-likelihood, as done by \cite{2017MNRAS.464.4789M}. This is achieved here by increasing the input errors (i.e. the measured RMS noise in the cube) by $(2N)^{1/4}$. 
This approach appears to yield physically credible formal uncertainties on the inferred parameters, whereas otherwise these uncertainties are unrealistically small. 

Within \textsc{GAStimator} we utilise a MCMC method with Gibbs sampling and adaptive stepping to explore the parameter space. The algorithm runs until convergence is achieved, and then the best chain is run (with a fixed step size) for 30,000 steps (with a 10\% burn-in) to produce our final posterior probability distribution. We then marginalise over the probability surfaces for each model parameter to identify a best-fitting value (the median of the marginalised posterior samples) and associated 68\% and 99\% confidence levels (CLs). 
A quantitative description of the likelihood of each model parameter is presented in Table 1.

The best-fitting disc model requires a non-zero scale height (71$\pm$16\,pc), and has a fairly large internal velocity dispersion (66 $\pm$ 5 km s$^{-1}$). This suggests a reasonably turbulent disc, with a $V_{\rm max}$/$\sigma\approx2.5$ (compared to V/$\sigma\approx10$ found in typical relaxed low-redshift discs; \citealt{2015ApJ...799..209W}). The major- and minor-axis PVDs extracted from the best fitting model are shown as black contours in the left and central panels of Figure \ref{fig:kinmsmodel}, overlaid on the data (orange scale). Overall the model is a good fit to the data outside of the anomalous emission region. 

In the right panel of Figure \ref{fig:kinmsmodel} we show the major-axis PVD created from the residual cube (data-model; orange scale), again overlaid with the best-fitting disc model (black contours). The major positive residuals are in the anomalous emission region, as expected, but residual emission is detected all the way to the systemic velocity. Additional, smaller (but still significant) residual features are detected at a similar velocity on the red-shifted side of the galaxy, and further out in the disc on the blue-shifted side. It is unclear if these features are in any way related to the dominant blue-shifted component, or are simply brighter molecular knots within the disc of the galaxy.

\subsubsection{Properties of the blue-shifted emission}

As Figure \ref{fig:kinmsmodel} makes clear, the anomalous blue-shifted emission we detect in NGC\,0708 cannot be explained by simple rotational motion of the ISM. To learn more we begin by isolating the gas in this anomalous component from that in the main gas disc. In Fig. \ref{fig:CM2spec} we constrain its spatial extent using the velocity dispersion peak seen in Fig. \ref{fig:708_mom01}, adopting the region $-0$\farcs$32<$ RA offset $<0$\arcsec\ and $-0$\farcs$52<$ Dec. offset $<-0$\farcs13 (indicated by the black box in Fig. \ref{fig:CM2spec}) relative to the peak of the 236 GHz continuum ($\mathrm{RA}=01^{\mathrm{h}}52^{\mathrm{m}}46$\fs46 and $\mathrm{Dec.}=+36$\degr09\arcmin06\farcs46). 
In velocity, we impose $v_{\mathrm{obs}}<4550$\kms, indicated in the PVD (Fig. \ref{fig:708_PVD}) by a grey dot-dashed line. 

We note that these cuts only include emission that is unambiguously outside the normal rotation of NGC\,0708. One could estimate the properties of the anomalous emission using the residual cube presented in Figure \ref{fig:kinmsmodel}, but it is unclear whether all the residual emission detected in the disc is related, or whether some might simply be due to brighter regions in the disc which are not well fitted by our axisymmetric model. Where appropriate below we discuss whether this choice would change our conclusions.

The right-hand panel of Fig. \ref{fig:CM2spec} shows the $^{12}$CO(2--1) integrated spectrum of the spatial region containing the anomalous emission, with the channels satisfying the adopted velocity criterion (i.e. the blue-shifted wing) highlighted in magenta. We also show the result of fitting two Gaussians to this spectrum, one for the anomalous gas (shown in blue) and another for the gas in regular rotation in the galaxy disc (shown in orange). The sum of these two Gaussians is shown in red. The spectrum of this region has a $^{12}$CO(2--1) integrated flux of $2.7\pm0.1$\,Jy\kms, while the flux associated with the anomalous emission only (the area under the blue Gaussian in the right panel of Fig. \ref{fig:CM2spec}) is $0.54\pm0.05$\,Jy\kms\ (both $1\sigma$\ statistical uncertainty). The anomalous emission is blue-shifted by 284 $\pm$ 6 km s$^{-1}$ from the galaxy systemic velocity, and has a line-of-sight velocity width (RMS/dispersion of the blue Gaussian: $\sigma_{\rm v,los}$) of 51 $\pm$ 5 km s$^{-1}$. The properties of the Gaussian fitted to the anomalous emission are listed in Table \ref{tab:bluegasprop}. We note that if we were to estimate the flux of all residual emission at this position, after subtraction of the disc model presented above, the flux of this anomalous emission would increase by a factor of $\approx$2.

Using our ALMA data we can estimate the extent of this anomalous emission in the plane-of-the-sky using the \textsc{imfit} task in \textsc{CASA}. In the channel where the fitted anomalous component reaches its peak intensity (4466 \kms) the emission is marginally spatially resolved, with a diameter of (0\farc28$\pm$0\farc07) $\times$ (0\farc20$\pm$0\farc05), or $\approx$66$\pm$16\,pc.

To estimate the amount of molecular gas associated with this feature, we need to assume a CO-to-H$_2$ conversion factor. However, the opacity and density of the gas in the anomalous feature is unknown, so we will conduct the analysis with three representative $\alpha_{\mathrm{CO}}$ values (following \citealt{Morganti2015}): a typical local/Milky Way factor ($\alpha_{\mathrm{CO,\,MW}}=4.6$\msun\,(K\kms)$^{-1}$\,pc$^{-2}$, as assumed for the bulk of the gas), a factor appropriate for the disturbed gas typically found in ultra-luminous infrared galaxies (ULIRGs; $\alpha_{\mathrm{CO,\,ULIRG}}=0.8$\msun\,(K\kms)$^{-1}$\,pc$^{-2}$) and a factor appropriate for optically thin gas ($\alpha_{\mathrm{CO,\,thin}}=0.34$\msun\,(K\kms)$^{-1}$\,pc$^{-2}$; see discussions of $\alpha_{\mathrm{CO}}$ in \citealt{Bolatto2013} and \citealt{Geach2014}). As above, we assume a line ratio {$T_{\mathrm{b,CO(2-1)}}/T_{\mathrm{b, CO(1-0)}}=0.25$ (\citealt{Edge2001})}. The derived masses (tabulated in Table \ref{tab:bluegasprop}) vary from $\approx10^6$ to $\approx10^7$\msun. This component thus contains only a small amount of the total molecular gas in this galaxy (<1\%). 

These derived masses enable us to derive plausible limits on the physical extent of the emitting component. We know that the anomalous gas is dense enough for CO molecules to survive and be excited enough to emit at their second energy level. Thus the volume density of this component is likely $\gtsimeq$1000 cm$^{-3}$. If the emission we observe is coming from a spherical cloud of uniform density then its radius must be $<$41\,pc (assuming $\alpha_{\mathrm{CO,\,MW}}$) or $<$17\,pc (assuming $\alpha_{\mathrm{CO,\,thin}}$). Assuming a cylindrical geometry (appropriate e.g. if this material is a filament aligned with our line-of-sight) and a diameter of 66\,pc (as measured above), the length of this filament must be $<$90\,pc (assuming $\alpha_{\mathrm{CO,\,MW}}$) or $<$6\,pc (assuming $\alpha_{\mathrm{CO,\,thin}}$). While this material is unlikely to have a constant density, these limits suggest that the anomalous emission is likely to arise from a physically-small component, that nonetheless has a large internal velocity gradient  and/or dispersion. 

\begin{table}
	\centering
	\caption{Properties of the blue-shifted anomalous emission.}
	\label{tab:bluegasprop}
	\begin{tabular}{lr}
		\hline 
		Property & Value\\
		 (1) & (2) \\
		\hline 
		Integrated flux & $0.54\pm0.05$\,Jy\kms\ \\
		Mass assuming $\alpha_{\mathrm{CO,\,thin}}$ & (1.5 $\pm$ 0.1)$\times10^{6}$\msun \\
		Mass assuming $\alpha_{\mathrm{CO,\,ULIRG}}$  & (3.6 $\pm$ 0.3)$\times10^{6}$\msun \\
		Mass assuming $\alpha_{\mathrm{CO,\,MW}}$ &  (2.1 $\pm$ 0.2)$\times10^{7}$\msun\\
		Velocity centroid shift ($V_{\rm los}$) & -284 $\pm$ 6 km s$^{-1}$\\
		Velocity width ($\sigma_{\rm v,los}$) & 51 $\pm$ 5 km s$^{-1}$\\
		\hline
	\end{tabular}
	\parbox[t]{0.45\textwidth}{{\textit{Note:} Mass, mean velocity shift (measured relative to the galaxy systemic velocity) and velocity dispersion (line broadening) of the blue-shifted anomalous emission identified in Figure \ref{fig:CM2spec}. Masses are estimated separately for an optically-thin $\alpha_{\mathrm{CO}}$, optically-thick ULIRG $\alpha_{\mathrm{CO}}$ and local/Milky Way $\alpha_{\mathrm{CO}}$. Statistical uncertainties are quoted at $1\sigma$.}}
\end{table}

\section{Discussion}
\label{sec:discussion}
In the above sections we have shown that NGC\,0708 hosts a rotating disc of molecular gas at its core, coincident with the dust-lanes visible in optical extinction (Figure \ref{fig:708_XRayJet}). It also contains an anomalous, kinematically-distinct blue-shifted component (shown clearly in Fig. \ref{fig:708_PVD}), which has a large internal velocity gradient/dispersion. The velocity field and PVD of NGC\,0708 indicate most of the gas is in regular rotation in the gravitational potential of the galaxy, although the disc is fairly turbulent. 
The blue-shifted feature, however, cannot arise from regularly-rotating material. The origin of this material is ambiguous, as both inflows and outflows can produce similar features. 
{Below we therefore consider potential evidence for inflow or outflow, using our ALMA (and eMERLIN) observations and literature data from other wavelengths.}

\subsection{Inflow}

Inflows are common in BCGs, especially those located inside cool-core clusters like Abell 262. Inflows can also be found in more normal galaxies, and can arise due to gas cooling or mergers.

In NGC\,0708 we can rule out the type of inflows seen in isolated and satellite galaxies. In such systems gas cooling tends to be slow and to occur primarily along the disc plane (see e.g. the simulations of \citealt{Agertz2009} and \citealt{Stewart2011}). It would then most prominently appear in the moment 1 map as gas in the outer disc that is not following the expected rotation pattern. However, the blue-shifted feature we see in NGC\,0708 is well-collimated, and essentially unresolved spatially, indicating that a significant component of the velocity is along the line of sight, i.e. \textit{out} of the plane of the disc. The blue-shifted feature is also very close to the centre of the galaxy.

Mergers tend to be identified most readily in optical images, by the disturbed morphology of the galaxy and tails of gas and stars, as gas inflowing on to the more massive galaxy tends to form tidal tails extending over many kiloparsecs. In contrast, \textit{HST} imaging of NGC\,0708 (Fig. \ref{fig:708_XRayJet}) shows no sign of a significant recent major merger, with no tidal tail or disturbance of the stellar body. The dust distribution is however messy and warped, suggesting that a minor, gas-rich merger may have occurred. However, the PVD feature detected by ALMA is well collimated and has a small extent in the plane-of-the-sky. Such a small, localised and central feature is unlikely to arise from a tidal tail. Therefore, although we cannot definitively rule it out, the collimation, size and position of the blue-shifted feature indicate that secular or merger-driven inflow is an unlikely explanation. 

This, however, leaves the most likely possibility. NGC0708 is a BCG which is expected to be fuelled by a quenched cooling flow\footnote{A `quenched' cooling flow is a system where cooling gas accumulates in the central galaxy slowly, because interaction of the AGN with the cooling flow stops rapid runaway cooling.}. Molecular gas clumps and filaments are expected to precipitate out of the hot medium and rain down onto the galaxy. In large samples of observed BCGs these in-falling clumps/filaments are common, multiple typically being found around each BCG or brightest group galaxy \citep[e.g.][]{Temi2018,2018ApJ...854..167G,2019MNRAS.489..349R,Olivares2019}. These in-falling filaments typically contain a large fraction of the total molecular gas mass of the BCG. 

Figure \ref{fig:coolingrate_mass} shows the ratio of the molecular gas mass contained in in-falling filaments to the total BCG molecular gas mass for a range of clusters, plotted against the classical cooling rate of the hot ICM \citep{2018ApJ...858...45M}, as presented in \cite{Olivares2019}. If the anomalous emission in NGC\,0708 comes from a single in-falling filament on the far side of the galaxy then we can place it on this diagram. The measurements from this paper for NGC\,0708 are shown in Figure \ref{fig:coolingrate_mass} with an orange star. We note here that we use the estimate of the molecular mass of this feature assuming a Galactic CO-to-H$_2$ conversion factor (to match the assumptions made in the literature sample), as listed in Table \ref{tab:bluegasprop}. If the true CO-to-H$_2$ conversion factor of the filament is lower, our measurement would move down. 

\begin{figure}
	\includegraphics[width=\columnwidth]{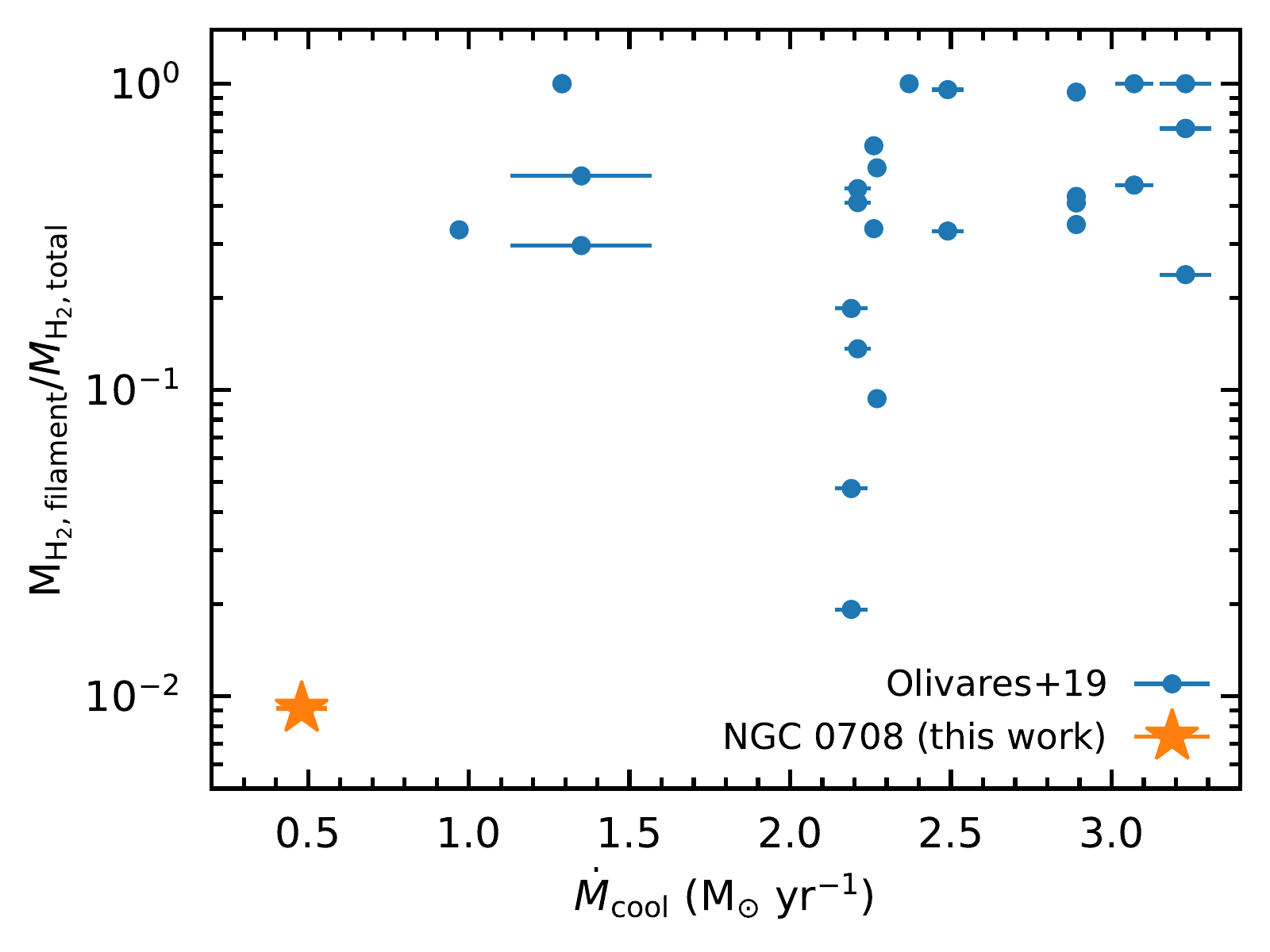}
	\caption{Ratio of the molecular gas mass contained within inflowing filaments to the total molecular gas mass of the BCG, plotted as a function of the classical cooling rate of the hot intracluster medium in each galaxy cluster \protect \cite{Olivares2019}. NGC\,0708 is shown as an orange star. It has a lower (classical) cooling rate than all other cluster objects (consistent with the lower total mass of Abell 262), and also contains a much lower fraction of its cold molecular gas in potentially inflowing filaments.}
	\label{fig:coolingrate_mass}
\end{figure}

In NGC\,0708 we only see clear evidence of a single non-equilibrium structure, that only contains a very small fraction of the cold gas in this galaxy. 
Only a single BCG from the \cite{Olivares2019} sample (RXJ1539.5) has a similar low-mass in-falling structure (the west filament, one of three detected in this system). 
However, detecting low-mass structures like that observed here requires high-sensitivity and high spatial-resolution observations, which are not available for all BCGs. 
In addition, Abell 262 (the parent cluster of NGC\,0708) has the lowest observed cooling rate in the \cite{Olivares2019} sample. It is thus possible that the low number of observed in-falling structures, and the low mass of the structure seen, could be typical of clusters with such low cooling rates.  

To match the observed velocity structure of this blue-shifted feature (which extends over $\gtsimeq$200 \kms\, while having a physical extension perpendicular to our line of sight of $\ltsimeq66$\,pc), it is clear that any such clump must have significant internal velocity structure. This is because (for any realistic potential) purely ballistic infall would require an extremely long column of gas (in which molecular gas would not survive) to create such a velocity gradient, while keeping the elongated gas structure perfectly aligned with our line of sight.

\begin{figure}
	\includegraphics[width=\columnwidth]{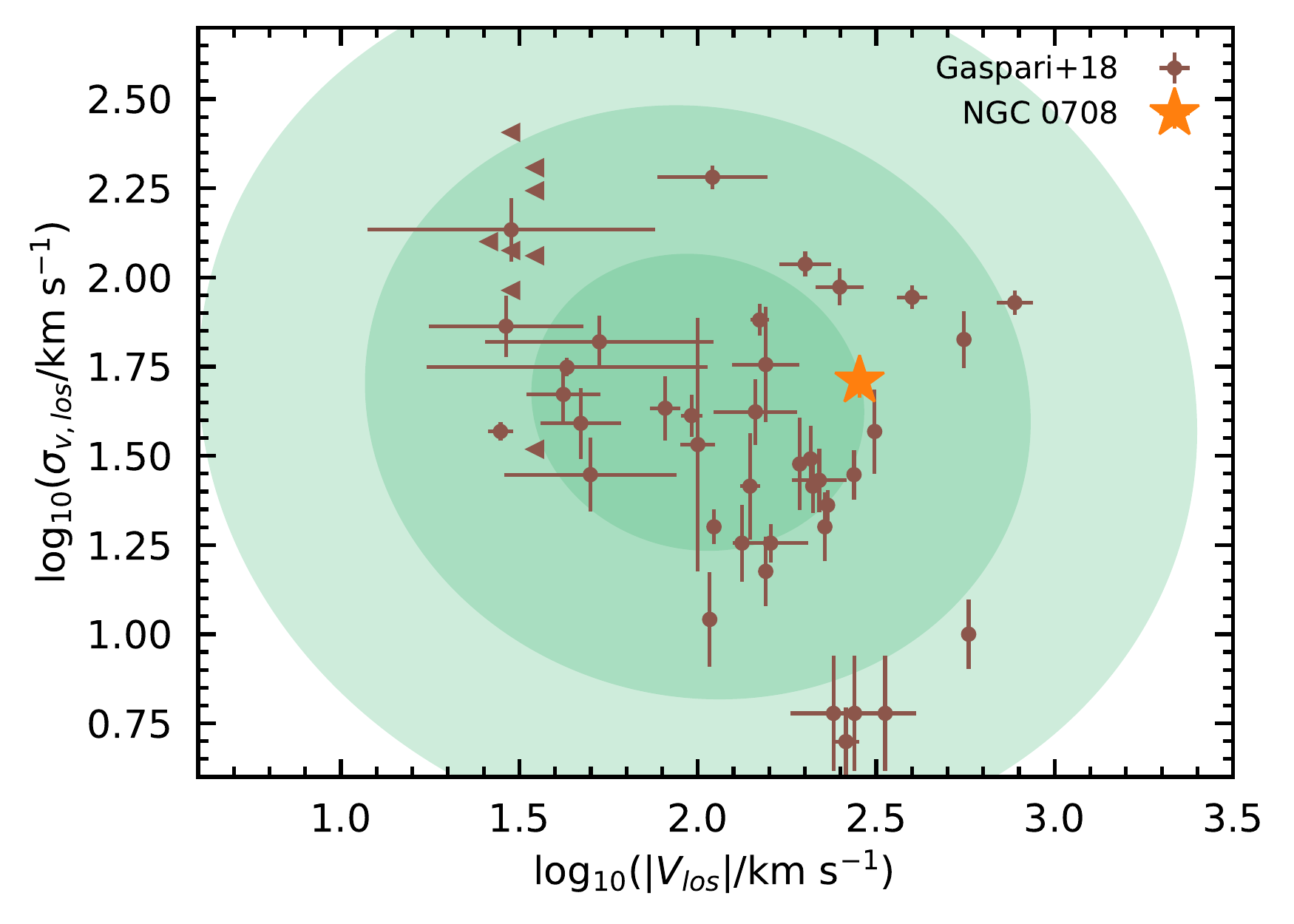}
	\caption{Velocity dispersion of molecular/ionised gas along individual lines of sight (`pencil beams') towards BCGs, plotted against the mean velocity shift of their line centroid from the host system's systemic velocity. The brown datapoints are reproduced from Table 2 of \protect \cite{2018ApJ...854..167G}, while our estimates from NGC\,0708 are shown as an orange star. Green contours show the $1$ to $3\sigma$ range in these quantities seen in the simulations of \protect \citealt{2018ApJ...854..167G}. The anomalous emission in NGC\,0708 has a significant internal velocity gradient, which is consistent at 1$\sigma$ with the expectation from CCA simulations for gas which has condensed out of the turbulent ICM. }
	\label{fig:gaspariplot}
\end{figure}

\cite{2018ApJ...854..167G} performed three-dimensional high-resolution hydrodynamical simulations of CCA and the related multiphase condensation cascade in group and cluster haloes, showing that the simulated clouds/filaments can have significant internal velocity dispersions, which match those observed in clusters and groups. In Figure \ref{fig:gaspariplot} we plot the velocity dispersion of the anomalous emission detected in NGC\,0708 against its mean velocity shift, and include for comparison the data (brown points; extracted from Table 2 of \citealt{2018ApJ...854..167G} and references therein) and simulations (green contours; 1, 2 and 3$\sigma$ confidence intervals of the emission detected along individual lines-of-sight) from \cite{2018ApJ...854..167G}. The anomalous emission we detect in NGC\,0708 is well within the scatter of the observational data, and deviates from the expectation of the simulations at only the $1\sigma$ level. 

Further evidence exists supporting a cooling-flow interpretation for this blue-shifted feature. For instance, the optical image of NGC\,0708 shown in Figure \ref{fig:708_XRayJet} shows that the dust on larger scales is disturbed, and thus there may be other filaments in the system, which are not (yet) traced by molecular gas (although dust is expected to be destroyed on very short timescales in the hot ICM, so its presence here would require fast dust formation mechanisms; \citealt{2010A&A...518L..50C}). 

The high velocity dispersion of the molecular gas disc of NGC\,0708 could be caused by material, such as the putative clump detected here, condensing out of the hot ICM and raining down on the disc. The turbulent Taylor number of this gas (defined as $V_{\rm rot}$/$\sigma$) is shown as a function of radius in Figure \ref{fig:taylorplot}. We estimate $V_{\rm rot}(R)$ using our best-fit parameters from Table \ref{fittable} in Equation \ref{v_as_func_r}, and $\sigma(R)$ by binning our observed moment two within elliptical apertures.  We note that formally these estimates of the turbulent Taylor number are lower limits, as observational effects (e.g. beam smearing, channelisation and the line-of-sight integration through the disc) preferentially lead us to overestimate the velocity dispersion. We highlight in grey the central regions of the galaxy where this effect is expected to be dominant. Outside of this region the turbulent Taylor number in the inner 100\,--\,300\,pc of NGC\,0708 is around 2.5, which suggests that one indeed expects the disc to be bombarded by condensing clouds and rare extended filaments \citep[see e.g.][]{2015A&A...579A..62G}. The outer regions have lower velocity dispersions,  and thus chaotic and filamentary condensation is less likely to be important there. We note, however, that this is not the only possible interpretation of the pervasive high dispersion of the molecular gas disc; see Section \ref{sec:outflowmodel}.

Overall we conclude that the blue-shifted material observed in NGC\,0708 could be a low-mass filament/cloud of material condensing out of the hot ICM and falling onto the core of the galaxy.

\begin{figure}
	\includegraphics[width=\columnwidth]{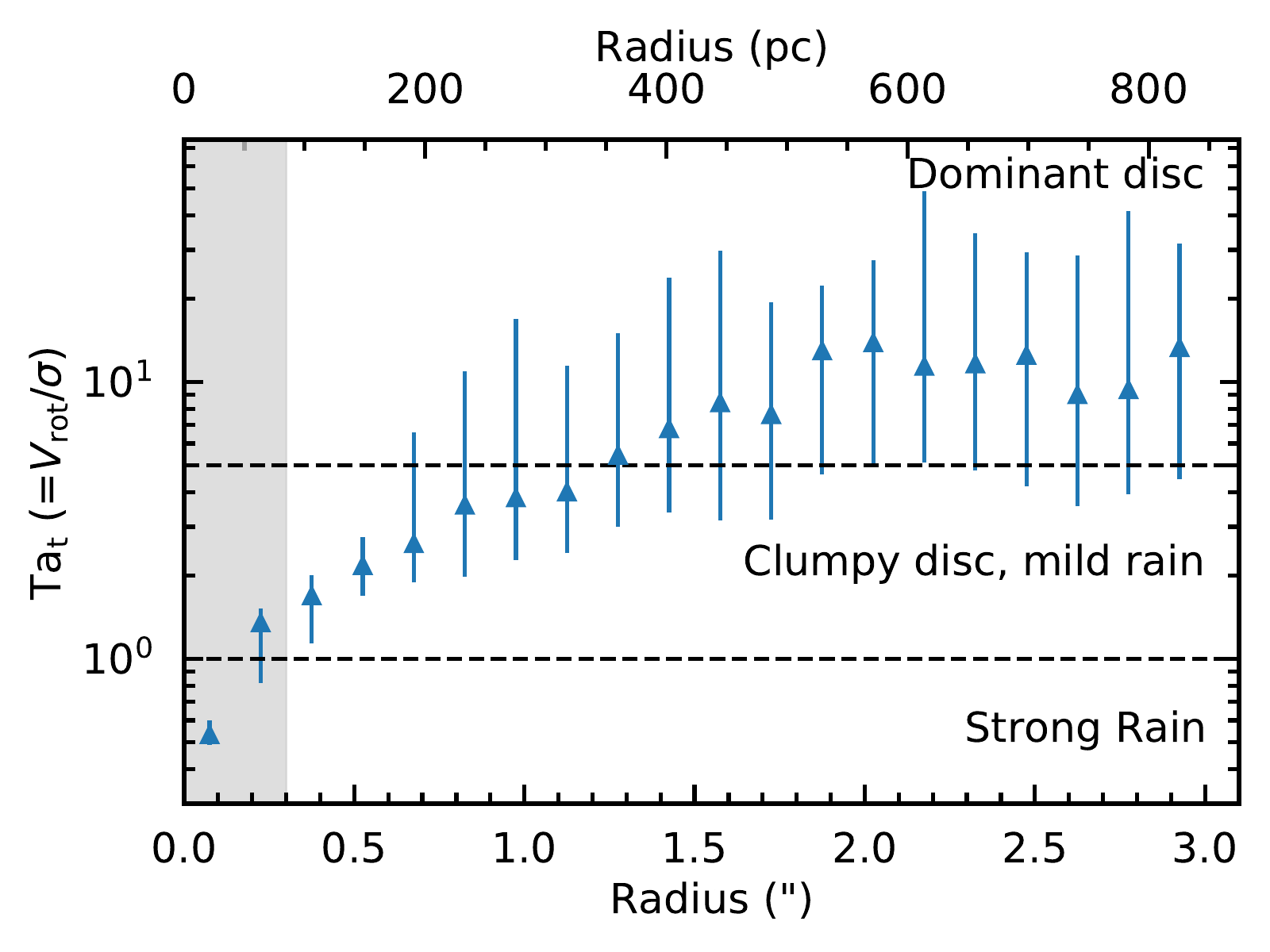}
\caption{Median turbulent Taylor number (Ta$_{\rm t}$) of the molecular gas disc in NGC\,0708 (estimated within elliptical apertures and shown as blue points with error bars showing the 16th to 84th percentiles of the distribution). Formally our measurements are lower limits, due to observational effects such as beam-smearing, and hence are plotted as upward facing arrows. The central regions, where these observational effects are expected to be dominant, are shaded in grey. Dashed black horizontal lines denote the expectation for gas arising from a strong CCA rain (Ta$_{\rm t}<1$), a mild rain (which forms a clumpy disc; 1$<$Ta$_{\rm t}<$5) and a system little affected by CCA rain  (Ta$_{\rm t}>5$) , as described in \protect \cite{2015A&A...579A..62G}. The central parts of NGC\,0708 have a low turbulent Taylor number, consistent with bombardment by a mild CCA rain.}
	\label{fig:taylorplot}
\end{figure}

\subsection{Outflow}
\label{sec:outflowmodel}
The other option for the origin of the blue-shifted feature is an outflow. Outflows can be caused by supernova- and/or AGN-driven winds as well as by jets directly impacting onto the ISM. 

Winds driven by supernovae tend to be large scale and are expected to be roughly bipolar, depending on the gas distribution around the star-forming region.
To lead to the feature shown in Fig. \ref{fig:708_PVD}, the wind would have to be very localised, or currently only interacting with a single (or at most a few) giant molecular cloud(s).  The total star formation rate (SFR) of NGC\,0708 is also very low, making this an unlikely scenario.

The blue-shifted anomalous emission is also offset from the AGN position as traced in 236 and 5\,GHz continuum emission ($\approx0$\farcs4 or $\approx113$\,pc offset; see the left-hand panel of Fig. \ref{fig:CM2spec}), making it unlikely that it is a quasar wind-driven outflow from the central AGN. 
The off-centre position of the feature could indicate a binary black hole system, with a dual AGN. However, neither radio nor X-ray observations detect accretion onto a second SMBH at this position, setting a stringent upper limit on the accretion power available to drive an outflow.  

AGN jets, on the other hand, are well collimated, strongly directional and can do work significantly away from the centres of galaxies, corresponding closely to the characteristics of the blue-shifted feature we observe. Jet-driven outflows do require the chance alignment of the jet with the ISM, but this is known to happen in a variety of sources (e.g. \citealt{Alatalo2011}; \citealt{Aalto2012}; \citealt{Morganti2015}; \citealt{Fernandez2020}). 

\begin{figure*}
	\includegraphics[height=8cm,trim=0cm 0cm 0cm 0cm,clip]{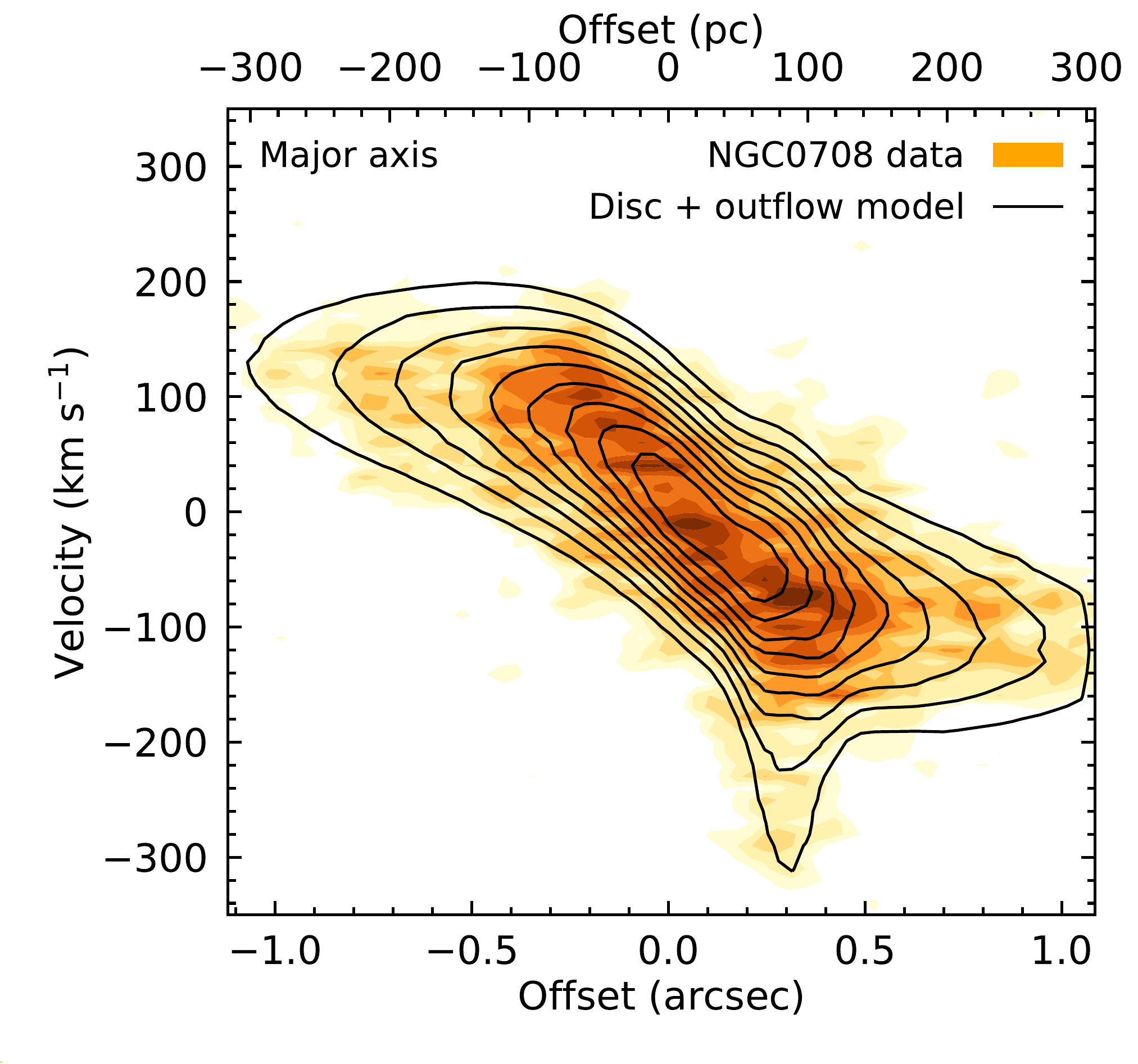}
	\includegraphics[height=8cm,trim=3.75cm 0cm 0cm 0cm,clip]{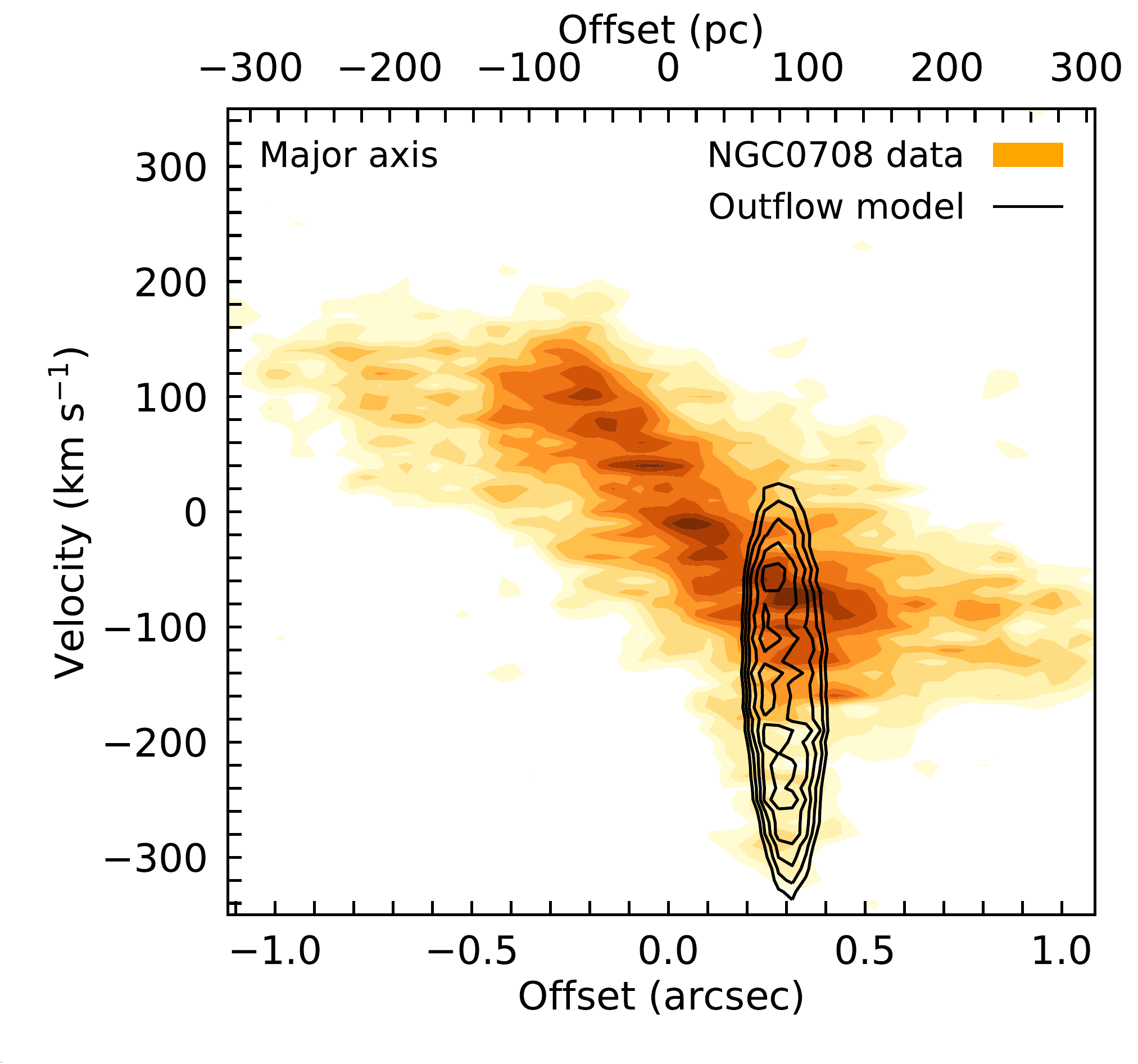}
	\caption{As Figure \ref{fig:kinmsmodel}, but adapting the \textsc{KinMS} model to include an expanding shell of material launched at an off centre position (see Section \ref{sec:outflowmodel}). In the left panel the full disc plus outflow model is displayed as contours overlaid on the observed data, while in the right panel only the outflowing material is displayed. Clearly, an expanding jet cocoon that drives an expanding shell of molecular gas can explain the anomalous, primarily blue-shifted emission in NGC\,0708, as the redshifted outflow is not easily identifiable in projection against the disc material.}  
	\label{fig:kinmsmodel_of}
\end{figure*}

As mentioned previously, an old AGN-driven jet is present in NGC\,0708 at 330\,MHz, but it is too large and not at the correct orientation to drive the putative outflow associated with the blue-shifted feature. 
We do detect signs of a jet at millimetre frequencies, pointing towards the putative outflow location (the Southern extension of the 236\,GHz emission discussed in Section \ref{sec:cocont}; see also Fig. \ref{fig:CM2spec}). However, we do not see signs of this jet at 5\,GHz suggesting that, if present, it must have a flat ratio spectrum (and is thus likely to be very young). The effects of shocks from such a small jet would be difficult to discern, as e.g. X-ray telescopes do not have the angular resolution necessary to separate emission at the putative hotspot from that of the nuclear point source. NGC\,0708 is known to have launched jets repeatedly within short timescales (with a repetition time of $\approx$28\,Myr), and at different position angles (likely due to precession of the central accretion disc). Indeed, cavities are detected in the hot ICM of Abell\,262 along the north-south direction \citep{Clarke2009}, suggesting the AGN axis was orientated in this direction in the past, and thus that a jet-driven explanation is plausible. 

The blue-shifted feature we find in NGC\,0708 is very similar to the jet-driven molecular gas outflow found in the Seyfert\,2 galaxy IC\,5063 by \citet{Morganti2015}, who observed a similarly jagged PVD with large deviations from the underlying large-scale rotation pattern at one specific off-centre position. \citet{Fernandez2020} also discovered a molecular gas outflow in the Seyfert\,2 galaxy ESO\,420\,G13, by detecting a velocity dispersion peak 440\,pc from the galaxy centre, again similar to the one detected here (see the right-hand panel of Fig. \ref{fig:708_mom01}). In these systems a young radio jet is driving into the molecular gas disc, inflating cocoons that drive the observed outflows, and causing the host galaxy discs to have high velocity dispersions (as also observed here; \citealt{Mukherjee2018}). While CCA and inflows are ubiquitous in BCGs, this suggests that an outflow explanation for this specific feature cannot be ruled out.

\subsubsection{Putative outflow geometry}
\label{kinmsoutflow_model}
Given the similarity between the anomalous blue-shifted emission we observe in NGC\,0708 and known molecular outflows, we consider the potential geometry such an outflow would have. 
Our observations only reveal a blue-shifted feature, while we would typically expect outflows to be (bi-)symmetrical. 
However, given that the launch velocity of the putative outflow is likely to be significantly away from the galaxy systemic velocity, the red-shifted side of the outflow is likely to be (at least partially) obscured by emission from the bulk of the rotating disc, as in some other similar sources \citep[e.g.][]{Fernandez2020}. To illustrate this, in Figure \ref{fig:kinmsmodel_of} we add an off-centre outflowing expanding sphere of gas (as in e.g. \citealt{Morganti2015}) to the kinematic model of a rotating disc presented in Section \ref{section:kinmsdiscmodel}. 

To model an expanding shell on top of this circular rotation model. we begin with the output `cloudlets' from \textsc{KinMS} (using the  \textsc{return\_clouds} option within the code). For every `cloudlet' within a (3D) radius $R_{\rm shell}$ of the assumed outflow launch point $P_{\rm launch}(x,y,z)$ we add a fixed velocity ($V_{\rm shell}$) directed radially away from $P_{\rm launch}(x,y,z)$. This creates an expanding bubble or shell of material, superimposed on the large-scale galaxy rotation pattern.  We include these five free parameters ($R_{\rm shell}$, $V_{\rm shell}$ and $P_{\rm launch}(x,y,z)$), and run an MCMC exploration of the parameter space as discussed in Section \ref{section:kinmsdiscmodel}. 
The best-fitting outflow position $P_{\rm launch}(x,y,z)$ is offset from the galactic centre by 0\farc00$^{+0.06} _{-0.05}$ in RA and -0\farc35$\pm$0\farc07 in Dec and 0\farc04$^{+0.03} _{-0.04}$ in the plane of the sky. The best fitting shell radius $R_{\rm shell}=$0\farc14$\pm$0\farc05 (or $\approx$ 33\,pc, fully consistent with the size estimated using \textsc{imfit} on the data directly in Section \ref{sec:bluecomp}). The velocity of the outflowing shell is estimated to be $V_{\rm shell}=$160$^{+14} _{-13}$ km\,s$^{-1}$, similar to that observed in known jet-driven outflows and simulations \citep[e.g.][]{Mukherjee2018}.
As shown in Figure \ref{fig:kinmsmodel_of}, such an expanding shell can well reproduce the kinematics of NGC0708, even the red/blue asymmetry.

\subsubsection{Putative outflow properties}

In this Section we calculate various properties of the putative outflow, and use these to discuss the viability of this explanation for the anomalous blue-shifted emission in NGC\,0708. 
We assume that the the outflow is a shell, where the outflow rate can be calculated as in \citet{Lutz2020}: 
\begin{equation}
\label{eq:Lutz2Mdot}
\dt{M}_{\mathrm{OF}} = \frac{M_{\mathrm{OF}}\, V_{\mathrm{shell}} }{\Delta R_{\rm OF}}
\end{equation}
where $\Delta R_{\mathrm{OF}}$ is the thickness of the shell, $M_{\mathrm{OF}}$ is the total outflow mass, and $V_{\mathrm{shell}}$ is the average velocity of the outflow.
We use here the masses estimated in Section \ref{sec:bluecomp} and listed in Table \ref{tab:bluegasprop} which (if this scenario is correct) correspond to gas contained within the blue-shifted outflow ($M_{\mathrm{OF}}$). However, we caution that the actual mass outflow rate may be underestimated, likely by a factor of $\approx2$ if the geometry presented in Section \ref{sec:outflowmodel} is correct (to account for the hidden red-shifted outflow). We conservatively adopt the estimate of $\Delta R_{\mathrm{out}}$ presented in Section \ref{kinmsoutflow_model}  (e.g. $\Delta R_{\mathrm{OF}}$ = $33\pm8$\,pc). In this case the outflow is a filled bubble, rather than a shell,  and mass outflow rates estimated will be lower limits if a thinner shell is present.
We assume $V_{\mathrm{shell}}=160^{+14} _{-13}$ km\,s$^{-1}$ as found in Section \ref{kinmsoutflow_model}.

To calculate the kinetic power of the outflow ($P_{\mathrm{kin, OF}}$), we use Equation 7 of \citet{Holt2006} rescaled to CO(2--1) from [\ion{O}{iii}] by \citet[see their Eq. 1]{Morganti2015}. Following both papers, we assume the relatively large line width of the outflowing gas reflects turbulent motion, so that the FWHM of the CO line represents the turbulent component of the outflow:
\begin{multline}
\label{eq:Holt}
\left(\frac{P_{\mathrm{kin,\,OF}}}{\mathrm{erg\,s^{-1}}}\right)=3.17\times10^{35}\left(\frac{\dt{M}_{\mathrm{OF}}}{\mathrm{M_{\odot}\,yr^{-1}}}\right)\\
\left[\left(\frac{V_{\mathrm{shell}}}{\mathrm{km\,s^{-1}}}\right)^{2}+0.18\left(\frac{v_{\mathrm{turb}}}{\mathrm{km\,s^{-1}}}\right)^{2}\right],
\end{multline}
where assume $v_{\mathrm{turb}}\approx\,\mathrm{FWHM}\approx\,2.355\,\sigma_{\rm v,los}$, and other terms are as defined previously. We note that including the turbulent term in this equation is controversial, but in our case this term is sub-dominant. We include it here to allow direct comparisons to other works.

The final property we calculate is the depletion time for the gas reservoir in NGC\,0708, assuming that the putative outflow continues at the same rate, via $\tau_{\rm dep} = M_{\rm tot}/\dt{M}_{\mathrm{OF}}$.

Table \ref{tab:OFK} lists the mass outflow rate, kinetic power and depletion times derived for each assumed $\alpha_{\mathrm{CO}}$ using the above equations.
We derive mass outflow rates between $\approx$7 and 104 M$_{\odot}$\,yr$^{-1}$, and kinetic powers between $\approx$10$^{41}$ and 10$^{43}$ erg s$^{-1}$. The entire molecular medium would be depleted by such an outflow in between $\approx$22 and 309 Myr. The assumption of a given  $\alpha_{\rm CO}$ is the dominant (systematic) uncertainty. We note that in other outflows the gas is typically (although not always) optically thin \citep{Lutz2020}, and thus we consider the lower power and outflow rate and longer depletion time estimates more likely.

\begin{table*}
	\centering
	\caption{Putative outflow properties.}
	\label{tab:OFK}
	\begin{tabular}{llrrrr}
		\hline 
		 \multicolumn{2}{l}{Property} & $\alpha_{\mathrm{CO,\,thin}}$ & $\alpha_{\mathrm{CO,\,ULIRG}}$ & $\alpha_{\mathrm{CO,\,MW}}$ \\
		 & & (1) & (2) & (3) \\
		 \hline
		Mass outflow rate & $\dt{M}_{\mathrm{OF}}$ (\msun\,yr$^{-1}$) &7.4 $\pm$  2.0 &  17.9 $\pm$  4.8 &  104 $\pm$  28\\
		\hline
		Kinetic power & $P_{\mathrm{kin,\,OF}}$ ($10^{40}$\,erg\,s$^{-1}$) & 6.0 $\pm$  1.8 &  14.5 $\pm$  4.3 &  84 $\pm$  25\\
		\hline
		Depletion time & $\tau_{\rm dep}$ (Myr) & 309 $\pm$  172&  129 $\pm$  73&  22 $\pm$  12\\	
		\hline
	\end{tabular}
	\parbox[t]{0.8\textwidth}{{\textit{Note:} Outflow mass outflow rate, kinetic power and depletion time for each of (1) optically-thin $\alpha_{\mathrm{CO}}$, (2) optically-thick ULIRG $\alpha_{\mathrm{CO}}$ and (3) local/Milky Way $\alpha_{\mathrm{CO}}$. Uncertainties are quoted at $1\sigma$.}}
\end{table*}

To ascertain if the jets launched by NGC\,0708 have enough energy to power the outflow, we calculate its jet power ($Q_{\mathrm{jet}}$). 
Equation 11 of \citet{Wu2009} converts the radio luminosity at 151\,MHz ($L_{151}$) to $Q_{\mathrm{jet}}$ (we follow their analysis and use a normalisation factor $f=10$). 
NGC\,0708 was observed as part of the 6th Cambridge (6C) survey at 151\,MHz at a resolution of $\approx7$\farcm$2\times7$\farcm2 ($\approx120\times120$\,kpc$^{2}$; \citealt{Baldwin1985}). The catalogue reports a continuum flux $L_{151}=0.78\pm0.075$\,Jy \citep{Hales1993}. 
The beam of these observations covers the whole of the old, large-scale jet in a spatially-unresolved manner, so we are forced to assume that an AGN restart would have produced a jet of similar power.
This assumption yields $Q_{\mathrm{jet}}=(1.32\pm0.01)\times10^{43}$\,erg\,s$^{-1}$, which would require a coupling factor with the ISM of 0.5\,--\,6\,percent (depending on $\alpha_{\mathrm{CO}}$ and the outflow geometry) to cause the outflow identified here. The efficiencies for an optically thin outflow are consistent with those seen in simulations (which yield jet-ISM energy transfer efficiencies of $\ltsimeq$1\,percent; e.g. \citealt{Nesvadba2010}; \citealt{Wagner2011}) and in known jet-driven molecular outflows \citep[e.g.][]{Alatalo2011,Alatalo2015,Davis2012,Aalto2012,Morganti2015,Fernandez2020}.

Overall we conclude that, although inflow onto the BCG is the most likely explanation for the blue-shifted anomolous feature seen in NGC\,0708, an jet-driver molecular outflow could also explain these observations. 

\subsection{SMBH mass}
The original goal of the observations presented here was to estimate the mass of the SMBH in NGC\,0708. 
As shown in Figs. \ref{fig:708_XRayJet} and \ref{fig:708_mom01}, the gas in the galaxy is disturbed (especially in the outer regions), and it does not lie in the equatorial plane of the galaxy, making this difficult. 
Despite this, the kinematics of the gas in the very centre of NGC\,0708 (around the AGN/continuum source seen; see Figs. \ref{fig:708_mom01} and \ref{fig:CM2spec}) seem fairly regular. Given the short dynamical times in this region, it is possible that this gas is sufficiently relaxed to allow us to constrain the central potential reasonably accurately.  

From the red-shifted side of the PVD, that appears fairly undisturbed by the outflow (see Fig. \ref{fig:708_PVD}),  we can make a crude estimate of the total mass enclosed within the inner resolution element of our data.
At 0\farcs088 ($\approx$24.8\,pc; one synthesised beam major axis from the galaxy centre) the projected rotational velocity is $\approx180$\kms, yielding an enclosed mass $M_{\rm enc}=2.15\times10^{8}$\msun\ (assuming pure rotation in a spherical potential, so $M_{\mathrm{enc}}=v^{2}(r)r/G$, where $v(r)$ is the de-projected rotational velocity at radius $r$ and $G$ is the gravitational constant). The total molecular gas mass within this radius is $\approx5.1\times10^{6}$\msun, and from the \textit{HST} F110W image we can estimate a stellar mass within the same radius of $\approx3\times10^{6}$\msun\ (assuming a very conservative F110W-band mass-to-light ratio of 2; see e.g. Fig. 11 of \citealt{Balogh2001}). This suggests a total dark mass of $\approx2\times10^{8}$\msun\ within 25\,pc of the centre of NGC\,0708. 

While this estimate is very uncertain due to the unknown degree of kinematic disturbance in the gas (and the approximate stellar mass-to-light ratio, absence of dust correction and standard CO-to-H$_{2}$ conversion factor), it is consistent with the SMBH mass estimated from the black hole mass -- central stellar velocity dispersion relations of \citet{Woo2002} and \citet{Donato2004}.

The Eddington luminosity of a black hole of this mass is $L_{\mathrm{Edd}}\approx2.67\times10^{46}$\,erg\,s$^{-1}$. In comparison, \citet{Clarke2009} estimated the total AGN kinetic luminosity to be $L_{\mathrm{AGN,\,kin}}=6.2\times10^{42}$\,erg\,s$^{-1}$ in NGC\,0708. 
As a percentage of the Eddington luminosity, this suggests the SMBH in NGC\,0708 is currently accreting at only $\approx0.023$\,percent of its Eddington rate. 

\section{Conclusions}
\label{sec:conc}

In this work we have presented $^{12}$CO(2--1) line and 236 continuum ALMA observations (along with 5\,GHz e-MERLIN continuum imaging), of the early-type galaxy NGC\,0708, the BCG in the galaxy cluster Abell\,262.
Our observations reveal a turbulent, rotating disc of molecular gas in the core of this galaxy. A marginally spatially-resolved `spike' of blue-shifted anomalous emission with a large line width is also present, approximately 100\,pc away from the galaxy nucleus. The (sub-)millimetre continuum emission peaks at the nucleus, but shows an extension towards this anomalous CO emission feature. No corresponding elongation is found on the same spatial scales at 5 GHz. 

The central kinematics of NGC\,0708 allow us to roughly estimate the non-luminous mass contained in the inner 25\,pc as $\approx2\times10^{8}$\msun, although given the somewhat disturbed kinematic state of the gas this estimate must be treated with caution. This value is, however, consistent with the SMBH mass expected for this source based on the black hole mass -- central stellar velocity dispersion relation.

We considered multiple potential causes for the anomalous blue-shifted emission detected in NGC\,0708, and conclude that two explanations are viable. X-ray observations of Abell 262, the host cluster of NGC\,0708, show that it is expected to have a turbulent and quenched cooling flow. Thus this feature could be a molecular filament that has condensed out of the hot ICM, and is falling onto the galaxy (consistent with predictions from CCA simulations). The fact that only a single such putative low-mass in-falling structure is detected in NGC0708 is consistent with it being located in a low-mass, low ICM cooling rate cluster.

While an inflowing filament explanation seems most likely, an alternative explanation is a molecular outflow, likely powered by a young, restarted radio jet. The kinematics of the anomalous emission in NGC\,0708 can be explained by an expanding shell/bubble of molecular gas, as seen in a variety of other galaxies with known jet-driven molecular outflows. A jet-driven scenario is also consistent with the observed episodic nature of the AGN in NGC\,0708, suggesting the outflow we are currently observing is young. This explanation would require the coincidence of the radio jet with the molecular disc, but the extension of the millimetre continuum provides tentative evidence of this, as long as the young jet has an inverted/flat synchrotron spectrum. 

To fully resolve the nature of this anomalous blue-shifted emission it would help to map the ionised gas at high spatial-resolution (e.g. with adaptive optics assisted integral-field spectroscopy), to see if the nebular emission in this system is consistent with cooling, or an outflow. Shock tracers (such as SiO) could also be observed with ALMA, to verify if a jet is impacting the molecular gas, and if so at what location. In either case, this source provides another intriguing place to study the interaction between cooling flows and AGN mechanical feedback.

\section*{Acknowledgements}

The authors thank the referee for comments which improved this work, and Grant Tremblay for advice during its preparation.
EVN and MDS acknowledge support from Science and Technology Facilities Council (STFC) PhD studentships. TAD acknowledges support from STFC grant ST/S00033X/1.
MB was supported by the STFC consolidated grants `Astrophysics at Oxford' ST/H002456/1 and ST/K00106X/1. M.G. acknowledges partial support by NASA Chandra GO8-19104X/GO9-20114X and HST GO-15890.020-A.
TGW acknowledges funding from the European Research Council (ERC) under the European Union's Horizon 2020 research and innovation programme (grant agreement No. 694343)

This paper makes use of the following ALMA data: ADS/JAO.ALMA\#2015.1.00598.S and ADS/JAO.ALMA\#2017.1.00391.S. ALMA is a partnership of ESO (representing its member states), NSF (USA) and NINS (Japan), together with NRC (Canada), NSC and ASIAA (Taiwan) and KASI (Republic of Korea), in cooperation with the Republic of Chile. The Joint ALMA Observatory is operated by ESO, AUI/NRAO and NAOJ.
This paper makes use of the following eMERLIN data: DD8014. eMERLIN is a National Facility operated by the University of Manchester at Jodrell Bank Observatory on behalf of STFC. This work has benefited from research funding from the European Community's sixth Framework Programme under RadioNet R113CT 2003 5058187. 
This paper also makes use of observations made with the NASA/ESA Hubble Space Telescope, and obtained from the Hubble Legacy Archive, which is a collaboration between the Space Telescope Science Institute (STScI/NASA), the Space Telescope European Coordinating Facility (ST-ECF/ESA) and the Canadian Astronomy Data Centre (CADC/NRC/CSA).

%%%%%%%%%%%%%%%%%%%%%%%%%%%%%%%%%%%%%%%%%%%%%%%%%%
\section*{Data Availability}
The data underlying this article are available in the ALMA archive, at \url{http://almascience.eso.org/aq/} projects ADS/JAO.ALMA\#2015.1.00598.S and ADS/JAO.ALMA\#2017.1.00391.S and the eMERLIN archive at \url{http://www.merlin.ac.uk/archive/archive_form.html} project DD8014. 
The datasets were derived from sources in the public domain.



\bibliographystyle{mnras}
\bibliography{NGC708_Papersv0} % if your bibtex file is called example.bib

%%%%%%%%%%%%%%%%%%%%%%%%%%%%%%%%%%%%%%%%%%%%%%%%%%

%%%%%%%%%%%%%%%%% APPENDICES %%%%%%%%%%%%%%%%%%%%%

%\appendix

%%%%%%%%%%%%%%%%%%%%%%%%%%%%%%%%%%%%%%%%%%%%%%%%%%

% Don't change these lines
\bsp	% typesetting comment
\label{lastpage}
\end{document}